# Halting SARS-CoV-2 by Targeting High-Contact Individuals

Gianluca Manzo*

*GEMASS*

*CNRS & Sorbonne University*

Arnout van de Rijt

*European University Institute*

*Utrecht University*

ABSTRACT

Network scientists have proposed that infectious diseases involving person-to-person transmission may be effectively halted by targeting interventions at a minority of highly connected individuals. Can this strategy be effective in combating a virus partly transmitted in close-range contact, as many believe SARS-CoV-2 to be? Effectiveness critically depends on high between-person variability in the number of close-range contacts. We analyze population survey data showing that indeed the distribution of close-range contacts across individuals is characterized by a small fraction of individuals reporting very high frequencies. Strikingly, we find that the average duration of contact is mostly invariant in the number of contacts, reinforcing the criticality of hubs. We simulate a population embedded in a network with empirically observed contact frequencies. Simulations show that targeting hubs robustly improves containment.

*Corresponding author: gianluca.manzo@cnrs.fr*

## 1. Introduction

Most policy measures that are currently used or considered to contain the novel coronavirus SARS-CoV-2 are aimed at broad groups of citizens (children, elderly, contact professions) or categories of meeting places (schools, restaurants, airports) (Zhang et al. 2020) and leave large chunks of the workforce idling or operating below capacity for extensive periods (Meidan et al. 2020). Such mass measures are widely viewed as necessary but they are costly. They have been shown to have a negative impact on national economic growth for several countries (Pichler et al. 2020: 17-20) through both forced industrial inactivity and consumer behavior change (see Goolsbee and Syverson 2020).

At the same time, a fair amount of evidence now suggests that the spread of many person-to-person viruses is driven by a small fraction of individuals, sometimes referred to as "superspreaders", who are responsible for the vast majority of secondary infections (James et al. 2007: fig. 1; Stein 2011; Wong et al. 2015; Sun et al. 2020). Many infected people appear to infect no one else. SARS-CoV-2 follows the same pattern. Estimates of the overdispersion parameter K —which, differently from population-level estimates of the basic reproductive number, $R_0$, quantifies heterogeneity across individuals in their capacity to generate secondary cases (Lloyd-Smith et al. 2005)—, consistently suggest that between 10% and 20% of cases are responsible for between 80% and 90% of secondary infections (Endo et al. 2020; Bi et al. 2020; Adam et al. 2020 Miller et al. 2020). Individuals generating an unusually high number of secondary infections are thought to have played a pivotal role in SARS-CoV-2 outbreak in many countries (for an overview, see Kay 2020; for a case study, see Hamner et al. 2020). This suggests that if one could identify and protect superspreaders, the virus may be controlled through focused interventions at lower overall cost.

The sources of high dispersion in individuals' capacity to generate secondary infections are not well-known. Some emphasize individual-level heterogeneity in infectiousness, such as differences in viral load, length of infection, and asymptomatic infection (Woolhouse et al. 1997; Galvani & May 2005; Cho et al. 2016). Others relate superspreading to specific contextual settings in which infectious individuals infect many others at once —so-called "superspreading events" (James et al. 2007; Hodcroft 2020). One way some suspect this may happen is that the buildings in which events take place facilitate airborne transmission by dispersing small droplets from any one source to many targets (Morawska and Milton 2020). Here we consider a third possibility, namely that the phenomenon of superspreading in SARS-CoV-2 has a network-structural basis. Some individuals may have jobs, living conditions, or social behaviors that generate many more close-range contacts than others. Their status as "hubs" in the network of close-range contacts could render them disproportionately instrumental in viral propagation, as they are both more likely to contract the virus, and once they have it, pass it on to more others. In some cases, these high degrees of contact derive from a specific role individuals play in an event, e.g. when a waitress or priest transmits a virus to many through serial dyadic contact. Without a consideration of network structure one may be inclined to blame the event and label it a superspreader event post hoc. Yet, an appreciation of the network structure of close-range interactions within these events would suggest a targeted policy protecting high-contact individuals, where undifferentiated event-level policies would have imposed high costs on large groups (Manzo 2020).

Theoretical studies have shown that when networks are characterized by high interpersonal variability in the number of contacts and thus the existence of hubs, epidemics may occur at a much lower per-contact transmission probability (Barrat, Barthélemy, & Vespignani, 2008, ch. 9). Under these circumstances, targeting hubs with transmission-reducing interventions (e.g., protective measures, behavioral restrictions, testing and quarantining if positive, treatment, and eventually vaccination) may effectively control viral spread in the population at large (Desző & Barabási 2001; Pastor-Satorras & Vespignani 2002).

The feasibility of this approach critically depends on the actual interpersonal variability in transmission-relevant contact. Early mathematical models of hub-targeting (Desző & Barabási 2001; Pastor-Satorras & Vespignani 2002; Cohen & Havlin 2003) as well as recent applications of this approach to SARS-CoV-2 (Hermann & Schwartz 2020) assume a scale-free spreading network, while empirical networks often deviate from this assumption (Jones & Handcock 2003; Clauset et al. 2009; Stumpf & Porter 2012; Broido and Clauset 2019). Nevertheless, degree-targeting may still be an effective strategy in the fight against SARS-CoV-2 if close-range contact exhibits high skew, with the majority of close-range contacts in society involving a small minority of individuals, as has been found for online contacts (Barabási & Albert 1999; Adamic & Huberman 2002; Vázquez et al. 2002) and sexual contacts (Liljeros et al. 2001; Trewick et al. 2013; Little et al. 2014). The approach to network intervention through preferential targeting of hubs has been variously elaborated over the years, both with respect to how measuring hubs (see, Kitsak et al. 2010; Montes et al., 2020) and how to reach them (see Rosenblatt et al. 2020); it has also recently been applied to SARS-CoV-2 (Hermann and Schwartz 2020). However, this literature overwhelmingly relies on observed or simulated networks that are of questionable relevance for the diffusion of a virus like SARS-CoV-2 for which direct close-range contacts aids droplet transmission (Mittal 2020). Studies based on short-range Bluetooth data, showing high interpersonal variability in the volume of face-to-face interactions (Mones et al. 2018; Sapiezynski et al. 2019), are promising but, for now, they only concern social encounters within small populations in single and specific social settings (like primary schools, hospitals, academic meetings or university) (see Cencetti et al. 2020).

The objective of this paper is to assess the effectiveness of hub targeting *versus* undifferentiated interventions for controlling SARS-CoV-2 spread in networks with empirically calibrated frequencies of close-range contact. For this reason, we draw on nationally representative datasets containing information on close-range contacts in various meeting locations, and the duration of each contact. As studies have shown that the spreading capacity of seeding hubs may be reduced when networks exhibit high clustering (see, in particular, Montes et al. 2020: figure 3, panel 3), we also aim to assess whether the effectiveness of hub targeting vis-à-vis undifferentiated intervention on networks with empirically-calibrated degree is stable across different network features for which lack of appropriate data impedes calibration.

The paper is organized as follows. From the survey data we derive degree distributions for close-range contact on a country scale (section 2). We then impose this empirical degree distribution on a synthetic social network with a tunable level of clustering (section 3.1). In this network, we introduce a virus with the main empirical features of SARS-CoV-2, and, by an agent-based implementation of a SEIR model, we let the virus spread through the network under various transmission conditions (section 3.2). We design different ways of reaching the best-connected nodes (section 3.3), and calculate how the trajectory of the epidemic varies under these interventions (section 4). From our simulation model we derive the hypothesis that interventions - such as vaccinations, medical testing, quarantining-if-positive, protections in high-risk professions, and informational campaigns - will be more effective when targeted at hubs rather than at random individuals (section 5). We conclude by discussing implications and limitations of the study (section 6)[1].

## 2. Data analysis

We draw on data from COMES-F, a survey conducted in 2012. In the survey a representative sample of about two thousand French residents report their close-range contacts (Béraud et al. 2015). The COMES-F survey data have several features that make it attractive for current purposes. First, compared with sensor-based digital data, representative survey data allow a comprehensive picture of contacts across social settings in the target population, thus generating a representative degree distribution. Second, among the major general population contact surveys conducted in Europe during the last decades (for a detailed comparative overview, see Hoang et al. 2019: table), COMES-F is the most recent, with the largest representative sample, based on paper diary, allowed respondents to report up to 40 contacts in their contact diaries, and, for each self-reported contact, recorded location, duration, and frequency. In addition, specific care was given to collect high-quality contact information for respondents aged 0-15. Finally, a recently conducted survey in six countries that we analyze in the Appendix (Belot et al. 2020) does not use the more thorough contact diary method, instead asking respondents for an estimate of the number of contacts.

Contact survey data are routinely employed by epidemiologists to build social contacts matrices, i.e. average contacts between age groups by places like school, public transportation, or home (Prem et al 2017). COMES-F have been frequently exploited in this way within compartmental models of SARS-CoV-2 spread in France (see, for instance, Di Domenico et al. 2020; Roux et al. 2020; Salje et al. 2020a; Walker et al. 2020). In contrast to prior use of the data, we rely on the entire cross-individual heterogeneity of the observed distribution of close-range contacts. We implement this distribution in a social network model of disease propagation (section 3), so that we can evaluate the effectiveness of interventions targeted at high-contact individuals (section 4).

The COMES-F survey was conducted in France during the first half of 2012. An initial sample of 24,250 was drawn from the French population excluding overseas territories through random-digit dialing of landline and mobile numbers. Using quotas for age, gender, days of the week and school holidays, 3,977 people who accepted to participate were sent a contact diary to complete. 2,033 (51%) contact diaries were returned (participants' age and household size were used as sampling weights to maintain representativeness). In these diaries, participants were asked to keep track of all short-range contacts over the course of 2 full days, and report on sex and age of these contacts, meeting context, and contact duration. Respondents were explicitly instructed to consider as a short-range contact someone they talked to at less than 2 meters, possibly including physical contact. To relieve the reporting burden, respondents were asked to record in the contact diary no more than 40 close-range contacts. For respondents aged less than 15, an adult member of the household completed the diary.

Specific questions concerned respondents currently in employment. In particular, they were asked whether they regarded their occupation as especially exposed to short-range contacts. This turned out to concern 257 respondents. These respondents had to indicate the average number of persons they estimated to meet every day because of their job. Should this number be higher than 20, those specific respondents were asked to enumerate only non-professional contacts when filling in the contact diary. Throughout the paper we will refer to diary-based contacts and job-related extra contacts respectively to differentiate the two types of measurement processes. This is an important distinction. It draws attention to one limitation of contact-diary based data collection: for the vast majority of respondents, contact diaries only allow to accurately estimate the total volume of close-range contacts but it is not possible precisely to distinguish the specific type of each of these contacts (family members, friends, co-workers, clients, unknown persons, and so on). In this respect, COMES-F is not different from other epidemiological diary-based contact surveys where only the place a contact occurs is recorded but not the precise nature of the contact (see, for instance, Mossong et al. 2008; Danon et al. 2013). Following previous analyses of COMES-F data (Béraud et al. 2015), we investigate separately diary-based contacts and job-related extra contacts, and explain later how we combine them to calibrate our simulations.

### 2.1 Contact volume and duration

The contact distributions are shown in figure 1. Through contact diaries, the 2,033 individuals reported a total of 19,728 per-day close-range contacts (left panel). The median number of contacts is 8 whereas the average is approximately 9.5. Respondents reporting a number of close-range contacts greater than twice (n=175) or even three times (n=36) the mean are not rare. The distributional character of the right skew is captured by the distribution of close-range contacts above 19 not significantly deviating at the 95% confidence level from a power law with a scale parameter 5.1 (n=175)[2]. Variance and skew are more pronounced among respondents who declared extra-job related contacts (n=257) (figure 1's right panel). Overall they reported 14,971 additional contacts. For those contacts the median is 30 whereas the average is approximately 58. The tail of the distribution in figure 1 above 17 does not significantly differ from a power law with a scale parameter 2.5 (n=190). The central tendencies of both distributions are consistent with those found in other diary-based contact surveys (see Hoang et al. 2019: 727-728). In both cases, averages are clearly driven by a small fraction of individuals reporting high numbers of short-range contact. The feature of high distributional skew is visible in recent smaller-scale contact surveys conducted in China (Zhang et al. 2019; Zhang et al. 2020). More particularly, the power law scale parameter estimates are similar to what was found for the UK Social Contact Study (Danon et al. 2012, 2013). In Appendix A1 we show that this variability persists within major demographic categories. Here we describe instead the relationship between the volume of contacts and their duration. As we treat high-contact individuals as leverage for effective intervention in viral diffusion dynamics, it is important to examine this relationship: the superspreading potential of hubs would be reduced if contacts were on average much shorter.

From a social network perspective, one may expect a negative correlation. As time and cognitive resources needed to sustain independent social relationships are limited (see, for instance, Dunbar 2016), individuals with many contacts may on average spend less time per contact. If this were the case then hubs may expose and be exposed by more people but per contact face less risk, reducing the criticality of hubs in the contagion. Face-to-face, close-range contacts may partially escape this logic, however. A dance instructor, for instance, who spends ten hours a day in a closed room giving private lessons to ten different small groups of four dancers during one hour. Such a respondent would typically declare to experience 40 contacts with skin touch per day for 1 hour, and would probably add to this some contacts at home for more than one hour a day. Thus this person would combine high contact frequency with high average contact duration. Large-scale surveys of social encounters indeed documented that these situations are frequent. It is common for people to be involved in different types of physically-closed social interactions —e.g. in family, friendship groups, classrooms, dance clubs, choir rehearsals, stadium visits, and manual team labor—, sequentially or simultaneously, at different time of the day, sometimes with more than one person at once (see, in particular, Danon et al. 2012: figs 1a and S2). When face-to-face interactions are at stake, individuals combine contact time across multiple and possibly simultaneous social interactions rather than experiencing them as independent and mutually exclusive events (like when one needs time and energy to build durable friendship or professional connections).

---

[2] The power law is fitted using the R implementation (in package igraph) of the maximum likelihood method developed by Clauset et al. (2009).

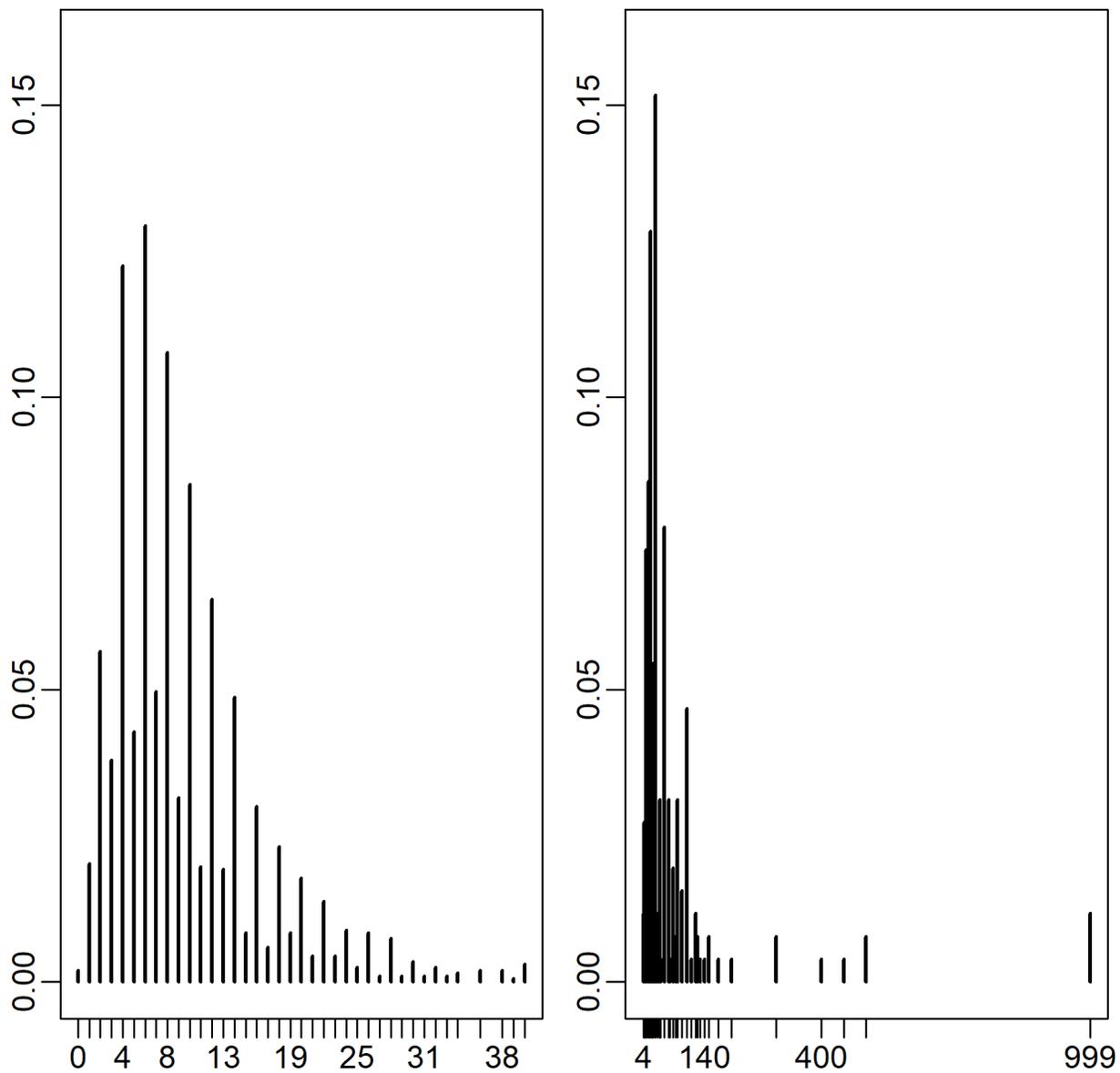

**Figure 1.** *Left*: Fraction of cases (y-axis) reporting a given number of close-range contacts (averaged over the two days) (x-axis) (n=2,033); *Right*: Fraction of cases (y-axis) reporting a given number of daily job-related contacts (x-axis) among respondents regarding their occupation as especially exposed to social contacts (n=257)

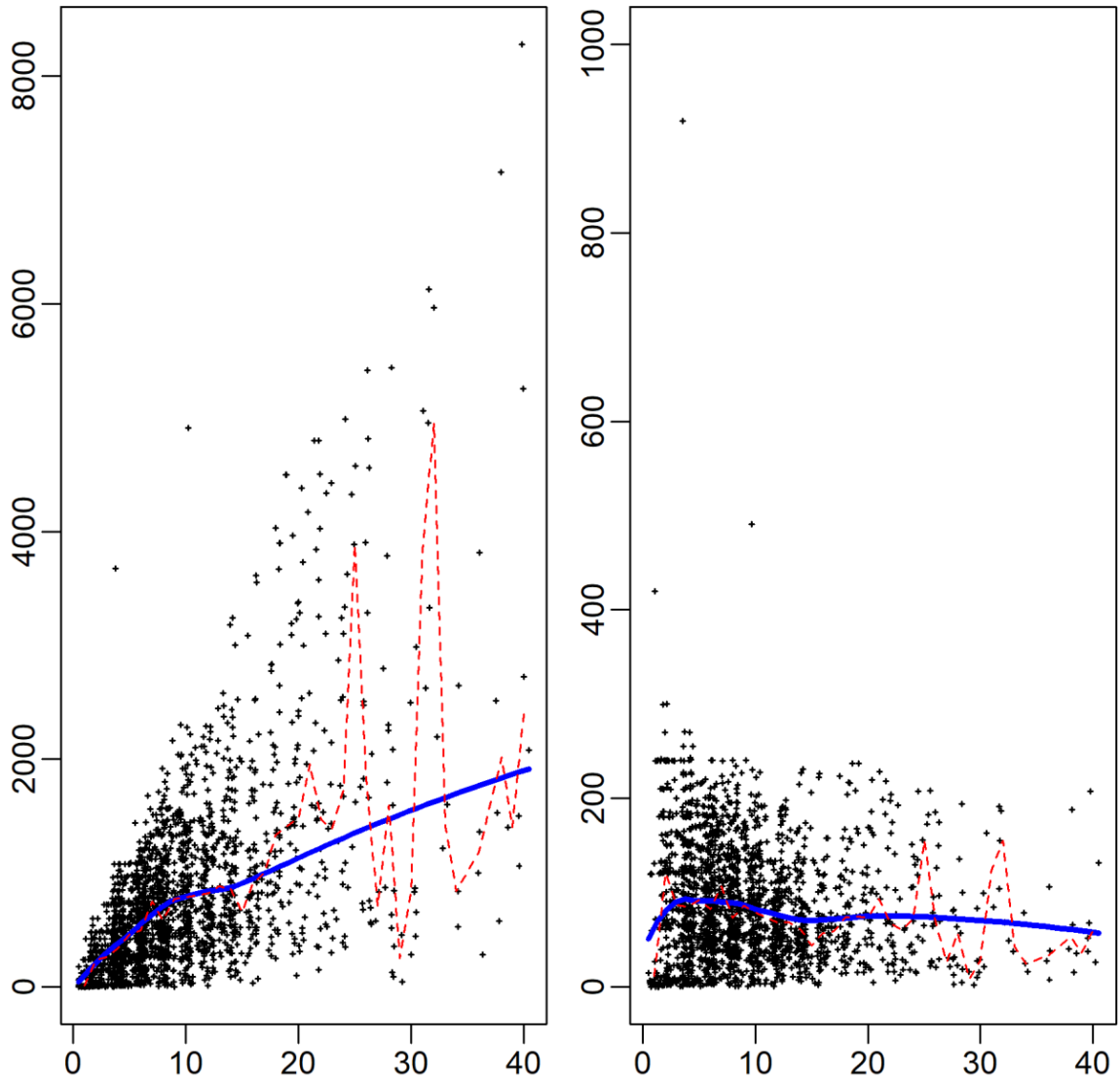

**Figure 2.** Left panel: Respondents' total daily contact duration (in minutes) (y-axis) as a function of daily # close-range contacts (x-axis). Right panel: Respondents' average contact duration (in minutes) (y-axis) as a function of daily # close-range contacts (x-axis). Points are jittered to avoid overlap. N=2029. Red dashed line: Median values of y-axis conditional on x-axis. Blue solid line: Local non-parametric regression curve (smoothing alpha parameter=0.5; polynomial degree=2) (fitted with R loess function). Total daily contact duration is computed as the sum (over all contacts) of the time the respondent declared having spent in each contact. Average contact duration is computed as total daily contact duration divided by daily # close-range contacts. Contact duration was recorded as a 5-category variable (1= < 5'; 2= 5'-15'; 3= 15'-60'; 4= 1h-4h; 5= > 4h): we consider the centroid of the interval (i.e. 2.5', 7.5', 22.5', 120', 240', respectively) to build the variables reported on the y-axis. Nota bene: Total daily duration may exceed 24 hours because many contacts happen simultaneously.

In the contact diaries, respondents additionally report for each contact its approximate duration. Figure 2 shows two scatterplots with on the y-axis the total duration of contact summed across all respondents' contacts (left panel) and the average duration of contact (right panel) by the number of reported contacts on the x-axis. For each plot we show the median y value for each x value and a LOESS curve. If time and cognitive resources would be limited for social encounters as they seem to be for more durable social relationships, one should observe a line with zero slope in the left panel and a sharply declining curve inversely proportional to x in the right panel. Strikingly, the actual empirical relationships are very different. In the left plot we find a monotonically increasing relationship with a slope that remains similar in magnitude over the observed interval between 0 and 40, a pattern that matches what was found by previous studies in the UK (see Danon et al. 2012: fig. S3c; Danon et al. 2013: fig. 2). Correspondingly, in figure 2's right plot we find little relationship between the number of close-range contacts and their average duration, with perhaps a slight decline in average contact length at high numbers of per-person contact. These results reinforce the criticality of hubs in spreading processes: The negative impact of higher numbers of contacts is not proportionally counteracted by brevity of contact. For this reason, in our simulations, the dyadic transmission probability does not depend on the total number of contacts that an agent has.

Unfortunately, COMES-F does not contain information on contact duration for respondents' self-reported estimates of job-related extra contacts. As a consequence, we cannot test the relationship between frequency and duration for the fraction of respondents reporting very high frequencies of contact (figure 1's right plot). For this reason, we adopt a conservative approach and base the calibration of the synthetic network underlying our main analysis on diary-based contacts only (see figure 1's right plot). In the appendix A3, however, we re-run all analyses under a combination of diary-based contacts and job-related extra contacts in such a way that a fixed transmission probability assumption is still defensible. Results are consistent and robust across these different specifications.

## 3. Model

Using COMES-F survey data we build an agent-based computational model in which the degree distribution of the synthetic network through which the virus diffuses is calibrated on the survey contact data (for other work using empirical network data in agent-based diffusion models, see Smith and Burrow 2018; Manzo et al. 2018). Our aim is to study the macroscopic consequences of cross-individual variability in close-range contact frequencies empirically observed at a country-level and assess whether this variability can be exploited for effective intervention in the ongoing epidemic. As such, we simulate a population the size of the COMES-F sample from which we eliminate four respondents who reported no close-range contacts.

### 3.1 Network construction and features

We connected 2,029 agents, each representing one respondent, according to two social network models. The first, which we will refer to as the "degree-calibrated" (DC) network, is the focus of our simulation. It is built to match the actual contact distribution and, at the same time, to tune network clustering, i.e. the propensity for two neighbors of a node to also be neighbors of one another. The second, which we will refer to as the "Erdős-Rényi" (ER) network, constitutes a benchmark against which to compare dynamics and effects of interventions.

In the DC network model, agents are first given a degree (number of network ties) precisely equal to the number of close contacts per day reported by each respondent in the contact diary (see figure 1's left plot). Then, to connect agents to one another, we adapted the configuration model, an algorithm that was proposed to generate random networks with arbitrary degree distributions (e.g. Jackson 2008: 83-85). To avoid duplicate links and self-links while ensuring an exact match between each virtual agent and an empirical respondent, we considered source agents in descending order of the to-be-generated degree rather than in random order, and then randomly picked available destination agents. Every time a connection was made, the degree of the two newly connected agents naturally increased by one. As soon as an agent reached the to-be-generated degree, it was excluded from the search algorithm. We found this procedure to always converge, achieving the intended empirical degree distribution.

However, the configuration model is known to be able to generate only a limited degree of clustering, which is constrained within this model by the imposed degree distribution and network size (Newman 2003: 202). Clustering is a crucial network feature that is known to attenuate the spreading capacity of high-degree nodes (see, for instance, Molina and Stone 2012). Recently, with particular reference to the COVID-19 crisis, Block et al. (2020) have presented simulation results showing that increasing local clustering of actors' ego-networks helps to mitigate the epidemic (see, in particular, box 2).

Thus, in order to assess the robustness of targeting hubs under various levels of clustering, we modified our generative network algorithm in a simple way. In particular, as soon as a given node reaches the required degree, before we exclude it from the search algorithm, we go through all its neighbors, and, we connect each of ego's neighbor pairs with probability $p$.

Our second network model is motivated by the fact that in most epidemiological models it is still common to assume random mixing according to which social contacts are assumed to happen at random within certain categories (Tolles and Luong 2020). From a network perspective, this amounts to postulating a random network where contact probabilities across individuals have little variability and the expected degree of each node is the average degree (Newman 2002; Barthélemy et al. 2005). We therefore also study an Erdős–Rényi (ER) random graph with the average degree observed in the survey (again, we only consider diary-based contacts), as a benchmark distribution. This random network is characterized by low variability in contact across agents and low clustering.

Table 1 shows network statistics computed over 100 realizations of the two networks. By construction, the DC network reproduces the features of the actual degree distribution. In particular, the mean is higher than the median, which suggests right-skewness. By contrast, the degree distribution of the ER network has essentially equal mean and median. The dispersion of the degree also strongly differs between the two networks, with the DC network exhibiting greater variation in the nodal number of links than in the ER network in the order of a doubling of the standard deviation (This difference is much larger when additional professional contacts are considered, see Appendix A3).

| | Average degree | Median degree | Stdev degree | Clustering coef | Deg-clust corr | Av path length | Diameter |
|---|---|---|---|---|---|---|---|
| *Degree-Calibrated* (DC) networks | | | | | | | |
| *p*=0 | 9.72 (0.00) | 8 (0.00) | 6.56 (0.00) | 0.01 (0.00) | -0.06 (0.01) | 3.47 (0.00) | 6 (0.00) |
| *p*=0.5 | 9.72 (0.00) | 8 (0.00) | 6.56 (0.00) | 0.43 (0.00) | -0.62 (0.01) | 4.38 (0.03) | 7.45 (0.50) |
| *p*=1 | 9.72 (0.00) | 8 (0.00) | 6.56 (0.00) | 0.57 (0.01) | -0.56 (0.01) | 5.52 (0.09) | 10.10 (0.59) |
| *Erdős-Rényi* (ER) network | | | | | | | |
| | 9.65 (0.50) | 9.72 (0.10) | 3.11 (0.10) | 0.00 (0.00) | 0.00 (0.03) | 3.60 (0.01) | 6.06 (0.24) |

**Table 1.** Topological features of the simulated contact networks (as a function of the clustering probability *p* for the DC network). Mean values across 100 network realizations (standard deviation in parentheses). Clustering coef=clustering coefficient; Deg-clust corr=Pearson correlation coefficient between nodes' degree and their clustering coefficient; Av path length=Average of the shortest path lengths; Diameter=Maximum of the shortest path lengths.

For the DC network, Table 1 also shows that the way we modified the configuration model efficiently generates increasing levels of clustering as the clustering probability *p* increases. As expected, when *p*=0, meaning that we do not force ego's neighbors to close triads, the DC network exhibits a very low level of clustering, essentially comparable to the ER network. By contrast, when *p*=1, meaning that we force the maximum number of links among a focal agent's neighbors, the level of clustering increases to 0.57, the maximum level we can reach given the structural constraints imposed by the actual degree distribution and the size of our synthetic population.

As COMES-F's contact-diaries do not contain information on possible close-range contacts among a respondents' contacts, we cannot empirically calibrate the clustering coefficient of the DC network. To the best of our knowledge, only Danon et al. (2012, 2013) in the UK designed contact-diaries to collect information from which clustering of close-range social encounters could be estimated. Their data show an average value around 0.46, with a considerable range of variation from approximately 0.07 to 0.7 depending on age category, meeting place, and distance from home (see, in particular, Danon et al. 2012: fig. 2b and 2c; 2013: figs 3 and 5). These estimations rely on complex rescaling procedures that may overestimate (up to a factor of 1.8) the true level of clustering (see, on this point, Danon et al. 2013: SI, § 5.3). For this reason, we opt for studying our model over the range of possible clustering levels generated by our algorithm. This range includes the 0.46 estimate.

The DC network is characterized by a negative correlation between the nodal degree and clustering, a correlation that becomes stronger as the overall level of clustering increases. This means that the higher the degree of a node the lower the fraction of ties among its neighbors. Thus high-contact nodes span across the network more than they cluster together. This pattern was found on real-world close-range contact networks in the UK (see Danon et al. 2012: fig. 2a). As discussed by Barabási (2014: 232-237), this negative correlation between nodal degree and clustering is the statistical signature of the presence of community structure within the network, a topological feature that makes hub-centered interventions especially effective: attacking the hubs means interrupting (or slowing down) communication among the modules (*ibid*: 236). By contrast, the correlation between nodal degree and clustering in the ER network is virtually nil.

Finally, when $p=0$, meaning zero probability of closing triads among a node's neighbors, in the DC network *average path length* and *diameter* are comparable to those of an Erdős–Rényi network with the same size and degree, consistent with what is usually observed in pure scale-free models (see Albert and Barabasi 2002:74). Highly connected individuals are effective in bringing many parts of the network together. As we increase clustering, the DC network's *average path length* and *diameter* increase, too, but remain low, achieving the combination of high clustering and reachability characteristic of small-world topologies (Watts & Strogatz 1998).

In sum, the DC network displays the actual long-tailed distribution of close-range contacts observed in representative data, while incorporating important topological features among which clustering, community structure, and short path length that previous studies have shown to be consequential for the spread of disease. Our goal is to assess whether our proposed mitigation strategy of targeting hubs robustly increases epidemic control in these degree-calibrated networks.

## 3.2 Agent-based SEIR model

We model disease propagation through the *Degree-Calibrated* (DC) and *Erdős-Rényi* (ER) networks by building a stochastic agent-based implementation of a SEIR model (Martcheva 2015). The SEIR model is a type of compartmental model that has been previously applied to the COVID-19 outbreak (Brethouwer et al. 2020; Kucharski et al. 2020; Li et al. 2020; Prem et al. 2020). In particular, we follow recent empirical parametrization (see Salje et al. 2020a) to determine how agents unidirectionally move from being (S)usceptible, to (E)xposed, (I)nfectious, and eventually (R)ecovered. Each iteration corresponds to one day. The time it takes for an agent to move from one state to the next is calibrated accordingly.

Upon infection agents first enter E where they stay 4 days; during this period, they are not infectious (for this value, see Salje et al. 2020a: 10). They then move to I where they become infectious, and can contaminate other agents over the course of 4 days (for this value, see Salje et al. 2020a: 10). Infected agents move to R with probability following a normal distribution with average 0.993 (and possible range at the agent-level between 0.990 and 0.996) (for the average value, see again Salje et al. 2020b) provided they have spent a number of days in I at least equal to a given recovery time. The recovery time follows a Poisson distribution centered on 2 weeks (with possible range at the agent-level between 1 and 6 weeks) (for these values, see empirical estimates in World Health Organization 2020: 14).

We combine this basic compartment structure with our network topologies such that the agents an infectious agent can infect are determined by the network of close-range contact (see Barrat et al. 2008: ch. 9). During each day, an infectious agent can only transmit the disease to its direct contacts. The dyadic, meaning agent-to-agent, transmission probability $r$ is assumed to be normally distributed with mean equal to 0.03, 0.05 or 0.07, and standard-deviation equal to 0.02[3].

To the best of our knowledge, currently there is no data that allows us to estimate the transmission probability at the dyadic level. For this reason, we follow a common procedure that simulates the model under different values of the likelihood of infection in order to assess whether the intervention of interest is robust across epidemics of different sizes (see, for instance, Block et al. 2020).

---

[3] Rather than limiting transmission over the short period of time where infectiousness seems to be highest, an alternative specification would consist in using a reversed-U shape (discrete or continuous) probability function over a larger infectious period. Hermann and Schwartz (2020: appendix), for instance, on the basis of epidemiological data concerning 94 Chinese patients, span dyadic transmission probabilities over a period of 14 days on a probability interval starting at 0.01, peaking at 0.3 (for days 5, 6, and 7), and progressively going back to 0.01. On a purely simulated network, and assuming perfect knowledge of the degree distribution, the authors target nodes with the highest degree, and show the effectiveness of this strategy to mitigate the virus spread. Thus, a modeling choice that requires making assumptions on a much larger number of values than ours leads to results that, as to the role of hubs, are in line with our own results.

In our model, the values of 0.03, 0.05 and 0.07 were chosen because they were able to trigger, on the DC network, epidemics where approximately 20%, 60% and 80% of agents were ever infected, thus allowing us to assess the effect of hub-targeted *versus* random interventions under very different scenarios[4].

All simulations start with five (randomly chosen) initially exposed infected agents. This is the lowest number of seeds that prevents excessive variability across simulation trials in our model population[5].

### 3.3 Interventions

We follow prior studies by considering targeted interventions that offer a set of agents protection against the virus (Pastor-Satorras & Vespignani 2002; Herrmann & Schwartz 2020): agents present in the Susceptible, Exposed, or Infectious compartment are moved to the Recovered compartment. The intervention thus prevents future infection of the targeted individuals, if susceptible, or prevents further spread from these targeted agents to other agents, if already infected. The intervention can represent any combination of measures, such as a vaccine against Covid-19, medical testing and quarantining-if-positive, protections in high-risk professions and targeted informational campaigns (Banerjee et al. 2020). We assume a government with a fixed daily (medical / technological / financial / ethical) capacity to intervene on $b$ individuals. This is implemented as follows: On day 1 (iteration 1), $b$ agents are selected from among all agents that are either in S, E, or I and moved to R, on day 2 (iteration 2) $b$ additional agents are selected who are currently in S, E, or I, and moved to R, and so on. We study four budgets for each intervention: $b = 1, 3, 5,$ and 10.

We consider three methods for selecting agents for intervention. The first method, "NO-TARGET", simply randomly samples $b$ agents for intervention each day, and is intended as a benchmark against which to contrast the other two methods. This method corresponds to what Pastor-Satorras & Vespignani (2002) refer to as "uniform intervention."

The second method, "CONTACT-TARGET", follows the strategy described in Cohen and Havlin (2010), whereby each day $b$ random agents are sampled who each select one random contact (without replacement) for intervention. Because of the friendship paradox (Feld 1991), these targets have above-average expected degree (Christakis & Fowler 2010). This is so because high-degree nodes are by definition overrepresented among other nodes' contacts (Feld 1991). The CONTACT-TARGET strategy is implementable in practice as a government could in fact randomly sample from the known population and have sampled individuals suggest their contacts. This is implemented as follows: On day 1, $b$ random agents are sampled from among all agents. For each sampled node a random network neighbor is sampled. The intervention is targeted at these b random neighbors. On day 2, again $b$ random nodes are sampled. For each sampled node a random network neighbor is sampled who had not previously been intervened on. The intervention is targeted at these $b$ random neighbors, and so on[6].

---

[4] In terms of basic reproductive $R_0$, if one computes this quantity for a network with heterogeneous degree (see, in particular, Olinky and Stone 2004: eq. 1), the chosen values of the dyadic transmission probability correspond, for the DC network (with no clustering), to virus spread characterized by $R_0$ respectively equal to approximately 1.52, 2.53, and 3.55, which allow to cover a wide range of $R_0$ (or $R_t$) values observed in different contexts, and/or at different time, in the current COVID-19 crisis (for France, see for instance, Salje et al. 2020b; Roux et al. 2020; in a comparative perspective, see Flaxman et al. 2020).

[5] In simulations on larger networks, which take much longer to run, in which we implemented the same degree distribution and used the same number of seeds (smaller fraction), we find that peaks naturally occur later, while the interventions we present next show qualitatively the same relative effects (results are available upon request).

The third method, "HUB-TARGET", assumes that agents' numbers of contacts are perfectly observed. During each iteration, nodes are targeted in strictly decreasing order of their network degree, starting with the $b$ largest hubs. This is implemented as follows: On day 1, the $b$ nodes with the $b$ highest degrees are selected and immunized; on day 2, the $b$ nodes with degree rank $b+1$ through $2b$ are targeted, and so on.

## 4. Results

Figure 3 shows the number of concurrently infected individuals over time for the four simulated networks when no interventions are taken. Results for the ER network are represented with dashed black curves and the three DC networks with low, medium and high clustering are shown as solid curves in respectively black, blue and red. Shaded areas represent variability in the inner 90% of simulated runs, that is, between percentiles 5 and 95. Panels A, B, and C present results for different dyadic transmission probabilities $r$, respectively 0.03, 0.05, and 0.07. Peaks are naturally higher at higher transmission probabilities, with vertical axes rescaled to accommodate these base differences across panels. In panel A peaks are not easily identified due to minimal spread.

A comparison between the ER and the DC network with virtually no clustering (Cc=0.01) shows that greater variability in degree generates higher peaks (for low dyadic transmission probabilities), and earlier and higher peaks (for middle and high dyadic transmission probabilities). This is consistent with theoretical results from formal models showing that in networks with high degree variance viral spread is faster, whatever the transmission probability is (Barthélemy et al. 2005). These results illustrate the impact of hubs: Highly connected individuals are more likely connected to the seeds and their neighbors. Once infected, they expose others early on, thus catalyzing viral diffusion. In the ER network, by contrast, there are no hubs to accelerate spread.

---

[6] The procedure is implemented in such a way that a/ if a randomly selected agent has no neighbor who had not been intervened on before, a new randomly selected agent is sampled as long as the condition is met, b/ the required number of agents to be intervened on is constantly adjusted as a function of the available population.

The epidemic size measured as the total number of ever infected agents is not easily seen in Figure 3. Table 2 shows these estimates along with 95% intervals. The DC network without clustering produces a larger epidemic than the ER network at low and medium dyadic transmission probabilities. At a high transmission probability, the ER network instead produces a larger epidemic. This pattern may be understood as follows: At high dyadic probabilities anyone is nearly guaranteed to eventually become infected except those with few ties. In DC networks there are many more agents with few ties than in an ER network. By contrast, at low transmission probabilities most agents are likely to escape the pandemic except hubs. In DC networks there are more hubs than in ER networks.

A comparison between the three DC networks in each panel of Figure 3 shows that epidemics are monotonically slower in networks with greater clustering. This is consistent with theoretical results from formal models showing that in networks with high degree variance increasing clustering attenuate hubs capacity to accelerate viral spread (Eguíluz and Klemm 2002; Serrano and Boguñá 2006). The mechanism is straight-forward: In networks with high levels of clustering, sources of infection are more likely to expose the same target rather than distinct targets, reducing overall exposure (see Molina and Stone 2012, fig. 3).

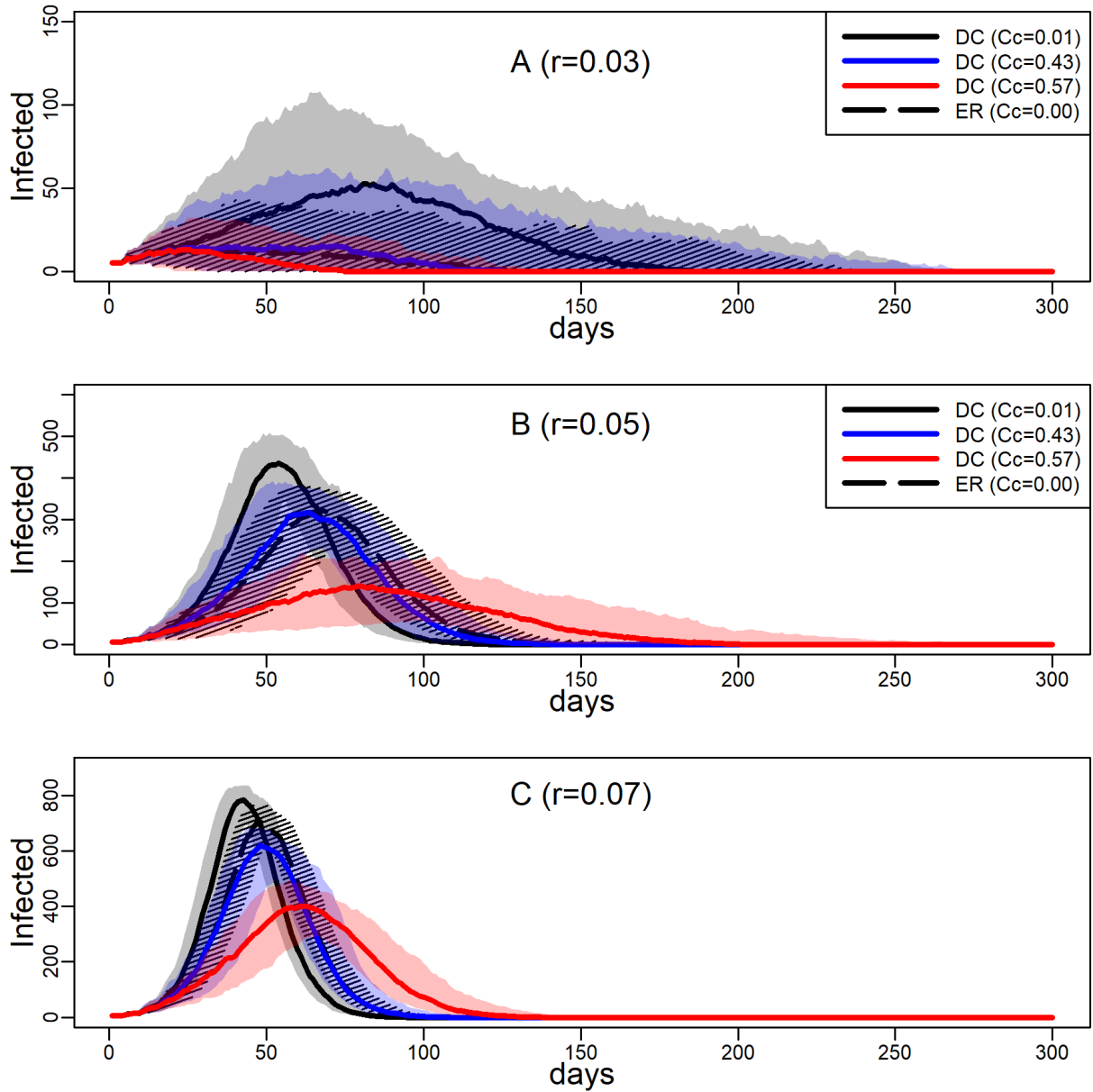

**Figure 3.** Number of infected agents (y-axis) by days (x-axis) (median of 100 replications) as a function of increasing values of the dyadic transmission probability *r* and clustering (see Legend). Lower and upper bounds of the shaded areas correspond to the 5[th] percentiles and 95[th] percentiles of the 100 replications. *n* = 2,029 agents. Solid line: *Degree-calibrated* (DC) networks; dashed line: *Erdős–Rényi* (ER) network with the same average degree.

| | $r = 0.03$ | | | $r = 0.05$ | | | $r = 0.07$ | | |
|---|---|---|---|---|---|---|---|---|---|
| | peak height | time | Epidemic size | peak height | time | Epidemic size | peak height | time | Epidemic size |
| DC (Cc=0.01) | **52.5** [0; 96] | 81 | 427.5 [13.9; 546.3] | **436.5** [307.75; 501.1] | 54 | 1241.5 [1181; 1312.05] | **783.5** [687.7; 836.15] | 43 | 1587 [1560; 1628.05] |
| DC (Cc=0.43) | **15.5** [0; 60.10] | 68 | 151 [7.95; 439.1] | **317** [218.1; 373.3] | 63 | 1173.5 [1108.85; 1248] | **619.5** [410.55; 684.45] | 48 | 1547.51 [509.95; 1586] |
| DC (Cc=0.57) | **13** [2; 26.15] | 19 | 46.5 [7.95; 158.65] | **140** [46.45; 210] | 79 | 996.5 [822.6; 1117.55] | **401.5** [286; 474.15] | 63 | 1510 [1414.75 1566] |
| ER | **13** [1.95; 30.05] | 24 | 102 [8.95; 313.6] | **322** [208.65; 376.35] | 68 | 1210 [1136.9; 1288.15] | **704.5** [584.05; 766.05] | 48 | 1661.5 [1619.95 1713.05] |

**Table 2**. Peak height (maximum # concurrently infected agents), time (in days), and epidemic size (# ever infected agents) on the *Degree-calibrated* (DC) networks with increasing clustering (Cc), and *Erdős–Rényi* (ER) network with the same average degree (rows) under low, middle and high dyadic transmission probability $r$ (column). Shown are median, 5% and 95% percentiles across 100 iterations.

Figure 4 shows the impact of the three intervention methods on viral diffusion in the DC network with low clustering, assuming an intermediate dyadic transmission probability. Peak reductions and timing are reported in table 4. Spread under intervention regimes is displayed as dashed curves in figure 4. Solid curves represent the no-intervention scenario, for contrast. Panel A shows results for the NO-TARGET procedure, whereby each day $b$ randomly selected susceptible, exposed or infectious agents are intervened on. The NO-TARGET procedure's maximally achievable impact, using the most generous budget considered, $b = 10$, corresponding to about 10% of the population being treated during the first 20 days, leaves the peak at 40% (Table 4: 174 / 436.5 ) of what it would have been without any intervention. Also, the peak occurs on about the same day.

Panels B show results for the CONTACT-TARGET procedure, which assumes that no global information on connectivity is available. Lacking this information, it attempts to find high-degree nodes by drawing a random sample of agents with unknown degree and selecting a random neighbor of each sampled agent for intervention. The figure shows this procedure is more effective than the NO-TARGET intervention. A budget of $b = 5$ CONTACT-TARGET interventions produces an impact comparable to a NO-TARGET intervention regime with $b = 10$ daily interventions. At $b = 10$, the CONTACT-TARGET intervention achieves a reduction down to only 14% of the peak in the no-intervention scenario (Table 4:

62 / 436.5). The peak is reached nine days earlier, after 45 days (CONTACT-TARGET) instead of 54 days (NO-TARGET).

The CONTACT-TARGET method would be more effective if randomly chosen agents would be able to select a random neighbor for intervention among relatively high degrees at a higher chance than network structure per se allows. Survey data suggest that targeting of certain professions may help effectively to identify high-degree agents (see appendix A1). To evaluate the maximally achievable impact of any degree-based intervention, panels C of figure 4 show the impact of the HUB-TARGET policy, whereby each day the $b$ previously untargeted agents with highest degree are targeted. A budget of 3 agents per day ($b = 3$) reduces the peak down to 30% (Table 2: 131.5 / 436.5). The peak occurs at the same time as without intervention. This reduction in peak daily infections achieved with $b = 3$ exceeds what NO-TARGET intervention achieves with ten agents per day ($b = 10$). With ten agents the pandemic is effectively prevented.

In sum, these results suggest that insights from formal models on abstract networks about the effectiveness of degree-targeting extend to networks with degree distributions that concord with contact survey data. And, by recalculating the figure 4 and table 3 for the Erdős–Rényi (ER) with the same average degree as the empirical degree distribution but lower degree variance (see table 1 above), it can be proved that it is precisely the skewness in the empirical distribution of close-range contact that makes hub targeting more effective than random targeting (see respectively appendix A2, figure A2a and table A2).

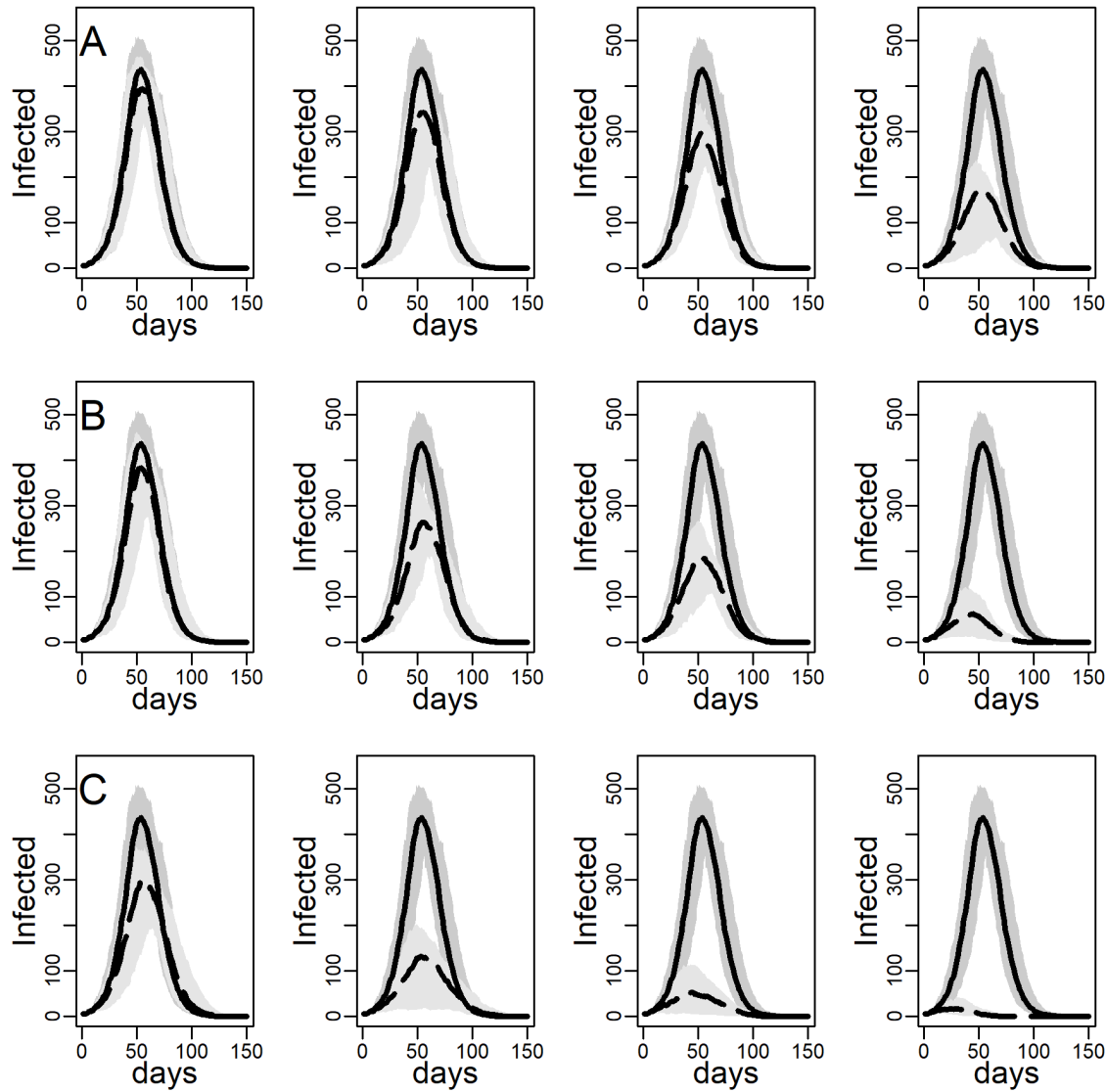

**Figure 4.** Number of infected agents (y-axis) by days (x-axis) (median of 100 replications) under three different interventions (rows) targeting 1, 3, 5, or 10 agents per day (columns). A – NO-TARGET intervention; B – CONTACT-TARGET intervention. C – HUB-TARGET intervention. Lower and upper bounds of the shaded areas correspond to the 5[th] percentiles and 95[th] percentiles of the 100 replications. Solid line: *Empirical-degree* (ED) network; dashed line: interventions. Dyadic transmission probability $r$=0.05 & Local clustering (Cc=0.01). $n$ = 2,029 agents.

| Degree-calibrated (DC) network | | | | | | | no intervention: H = 436.5 [307.75; 501.1]; T = 54 | |
|---|---|---|---|---|---|---|---|---|
| | | *b* = 1 | | *b* = 3 | | *b* = 5 | | *b* = 10 |
| | peak height | time | peak height | time | peak height | time | peak height | time |
| NO-TARGET | 394 [306.5; 462.1] | 55 | 341.5 [144.55; 407.35] | 55 | 295 [196.45; 362] | 52 | 174 [45.8; 226.3] | 52-53 |
| CONTACT-TARGET | 382.5 [222.85; 447.65] | 54 | 262.5 [131.95; 338.00] | 54-57 | 184 [84.5; 255.15] | 54 | 62 [11.85; 112.2] | 45 |
| HUB-TARGET | 298.5 [152. 364.2; 182] | 56 | 131.5 [15.6; 192.95] | 53 | 52.5 [3.95; 112.45] | 44 | 16 [3.95; 35.05] | 23-32 |

**Table 3**. Peak height (maximum # concurrently infected agents) and time (in days) under three interventions (rows) and four budgets (column) on the *Degree-calibrated* (DC) network. Dyadic transmission probability *r*=0.05 & Local clustering (Cc=0.01). Shown are median, 5% and 95% percentiles across 100 iterations.

Figures 5 and 6 explore the robustness of this result under increasing levels of clustering in the DC networks. Panels A, B, and C again represent results for the three intervention methods separately, again assuming an intermediate dyadic transmission probability (r=0.05). We consistently find a substantial improvement in virus-spread control of the contact- and hub-targeting methods over the random targeting method. In the Appendix A2 we further explore the robustness of our results under different assumptions on transmission probabilities and clustering (see in particular, figs. A2b-f). Results are qualitatively unchanged.

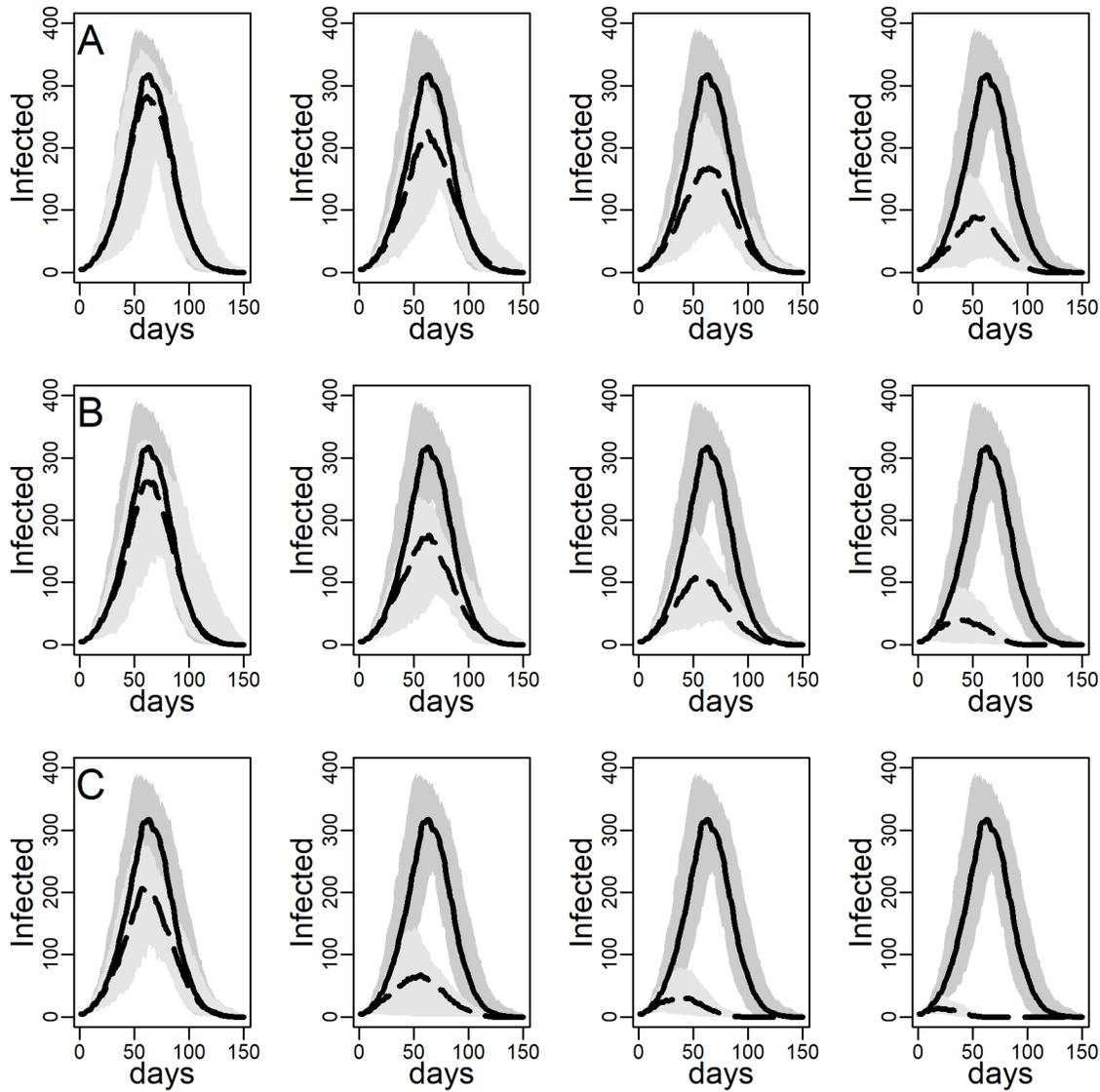

**Figure 5.** Number of infected agents (y-axis) by days (x-axis) (median of 100 replications) under three different interventions (rows) targeting 1, 3, 5, or 10 agents per day (columns). A – NO-TARGET intervention; B – CONTACT-TARGET intervention. C – HUB-TARGET intervention. Lower and upper bounds of the shaded areas correspond to the 5[th] percentiles and 95[th] percentiles of the 100 replications. Solid line: *Degree-calibrated* (DC) network; dashed line: interventions. Dyadic transmission probability *r*=0.05 & Local clustering (Cc=0.43). *n* = 2,029 agents.

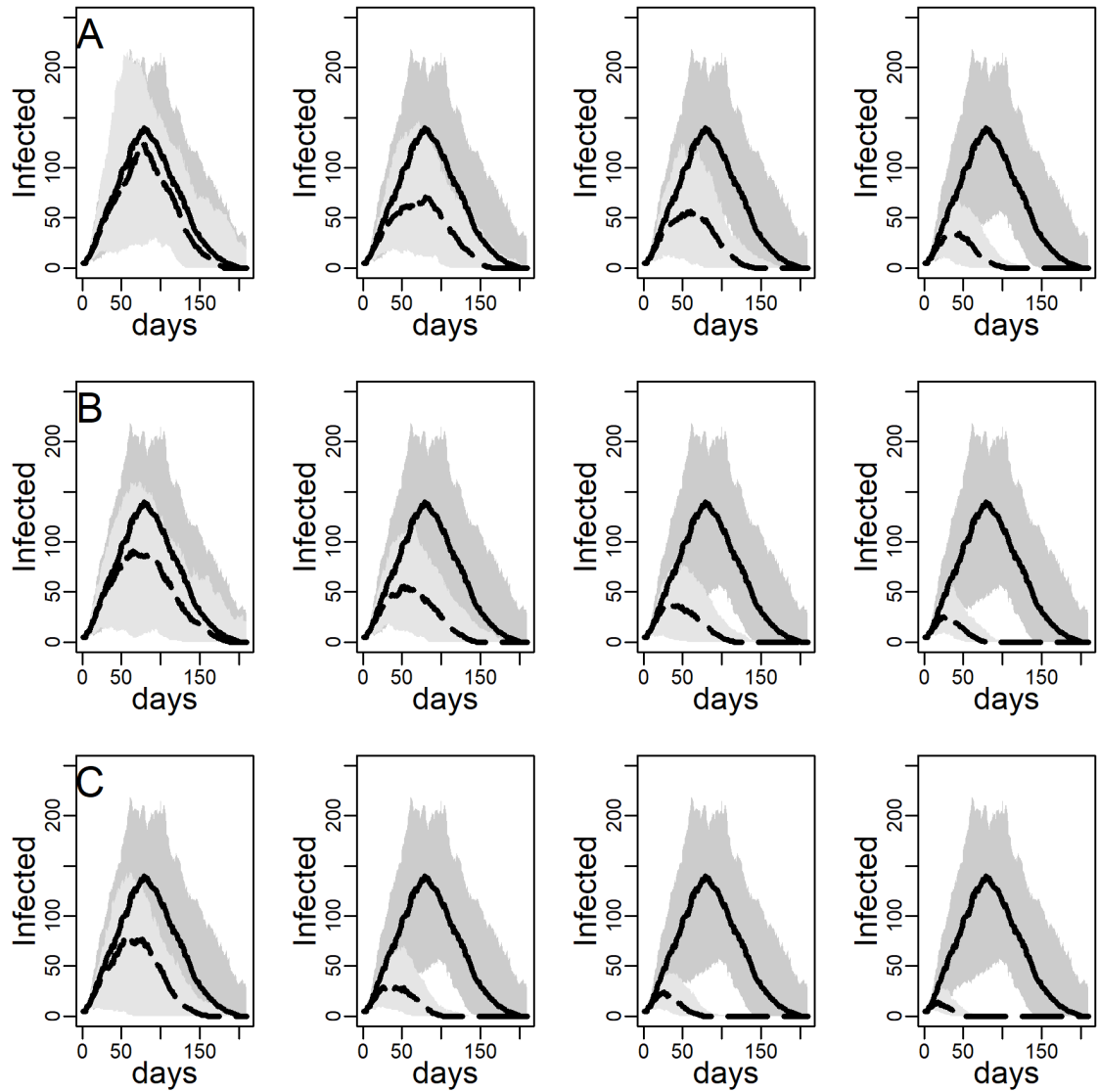

**Figure 6.** Number of infected agents (y-axis) by days (x-axis) (median of 100 replications) under three different interventions (rows) targeting 1, 3, 5, or 10 agents per day (columns). A – NO-TARGET intervention; B – CONTACT-TARGET intervention. C – HUB-TARGET intervention. Lower and upper bounds of the shaded areas correspond to the 5[th] percentiles and 95[th] percentiles of the 100 replications. Solid line: *Degree-calibrated* (DC) network; dashed line: interventions. Dyadic transmission probability *r*=0.05 & Local clustering (Cc=0.57). *n* = 2,029 agents.

## 5. Discussion

In the absence of a vaccine, countries worldwide seek to contain viral spread through a combination of social distancing, protective measures, informational campaigns, testing, and contact tracing (Sustained Suppression, *Nature Biomedical Engineering*, 2020). Yet there are clear limits on medical, technological, and financial resources and on the ability to durably restrict individual mobility, raising the question of how to prioritize. Our results suggest that all these interventions will generally be more efficient when targeted at individuals suspected or known to have close-range contact with many others. Once an effective vaccine has been developed, it may remain available in small quantities only for some time and/or face skepticism by large fractions of the population (Peretti-Watel et al. 2020). Based on our simulation results we can expect vaccination to reduce spread to a greater degree when high-contact individuals are given the first vaccines.

How could public policy effectively try and identify high-contact individuals? We propose two possible methods. First, the approach we systematically studied in our simulations is agnostic of who the high-degree individuals are and targets random acquaintances of random individuals, who statistically have high expected degree. This method was found effective in detecting past flu outbreaks (Christakis & Fowler 2010), and robust against missing network data (Rosenblatt et al. 2020). In our simulations this method shows to be reasonably effective yet at the same time is conservative in assuming no knowledge of degree or use thereof. If individuals did not nominate random contacts but instead those they know to have many other contacts, the difference targeted intervention could make would be greater.

The second method we suggest would exploit the covariation that seems to exist between individuals' occupation and the volume of their daily close-range contacts. Previous large-scale surveys on face-to-face encounters have documented that some professions (like teachers, service workers or health care workers) are especially exposed to close-range contacts (see Danon et al. 2013: fig. 4). We replicate this result in a survey on six different countries: Some professions involve ten times as many close-range contacts than others, with elementary school teachers, cashiers, order clerks, retail salespersons, and administrators topping the list (see Appendix A1, table A1).

The correlation between profession and contact frequency could be exploited in two different ways. One may want to directly target workers in professions characterized by high frequencies of contact. This approach would have the advantage of making it possible to set the interventions on the basis of category-based institutional lines, thus avoiding potential privacy and discrimination issues associated with targeting individuals. For instance, the occupational categories used in the international comparative survey we exploited follow a common international standard used by the US Bureau of Labor Statistics. Preferential protective legislation could be set on its basis. On the other side, however, this approach ignores the considerable within-occupation variability in workers' exposure to social contacts that we also documented (see Appendix A1, figure A1b). Undifferentiated category-based interventions, by protecting individuals within the category who are below the average exposure, waste resources. Another way to exploit the covariation between individuals' occupation and the volume of their daily close-range contacts would thus be to inject this information within the method of targeting random acquaintances of random individuals. According to this hybrid approach, randomly sampled individuals may be asked to preferentially report random social contacts within a given list of highly socially exposed professions. These contacts would then have a higher expected degree, rendering the method more effective.

One may expect this approach especially to benefit low-income workers. Quasi-experimental evidence suggests that the substantial income gradient in the impact of the pandemic on mortality is strongly mediated by low-income workers being trapped in low-paid jobs with high exposure to social contacts (Brandily et al. 2020). A hybrid strategy that combines occupational exposure with random acquaintances of random individuals to identify high-contact individuals to be protected/tested preferentially may thus be highly effective in reducing the overall death toll associated with SARS-CoV-2.

## 6. Conclusion

In this paper our goal was to assess the effectiveness of preferentially targeting hubs *versus* undifferentiated interventions for controlling SARS-CoV-2 spread. With this aim in mind, we moved away from the standard compartmental models that rely on random mixing assumptions toward a network-based modeling framework that can accommodate person-to-person differences in both infection risk and ability to infect others stemming from differential connectedness. Differently from past studies, we simulated virtual epidemics on networks with empirically calibrated frequencies of close-range contact. This framework allowed us to model rather than average out the high variability of close-contact frequencies across individuals observed in contact survey data. Results of simulations calibrated with empirical close-range contact distributions exhibiting right skew show large improvements in epidemic containment when shifting from general to targeted interventions. The relative effectiveness of preferentially targeting hubs proved highly robust across changes in degree skewness, clustering, and infection probability, as well as across epidemics of various sizes.

Our study has several limitations. First, the recommendation of prioritization of hubs in interventions is based on an assessment of effects on overall containment. There may be reasons to prioritize differently, e.g. protecting those in the medical profession dealing with SARS-CoV-2 as to maintain maximum capacity to treat. Alternatively, the protection of highly vulnerable subpopulations may reduce the overall death toll. The present paper does not speak to these alternative considerations as medical capacity and death rates are not modelled.

Second, our model is silent on the specific content of the actions to be performed on each finally selected individual. We only provide a method to maximize the efficiency of that selection. In the absence of a vaccine, moving from the model to the real-world world, an intervention could involve a combination of: (a) testing and quarantining-if-positive, (b) additional provision and mandation of protective instruments such as face masks and transparent physical barriers, (c) closer monitoring and tracking with mobile devices, and (d) targeted and contextualized informational messages stressing the importance of certain acts of social distancing and use of protective measures. Targeted messages can be relatively inexpensive as they are performed at distance (Marcus 2020) and evidence suggests they have strong health-behavioral effects (Noar et al. 2007), including recent field-experimental evidence to this effect for SARS-CoV-2 (Banerjee et al. 2020).

Finally, a factor that could limit the superspreader status of hubs and the effectiveness of hub targeting is contact time, namely if contact time were inversely proportional to the number of contacts. In this case hubs' shorter average per-tie duration of contact may be associated with lower risks of contracting and spreading the coronavirus (provided the probability of transmission is negatively correlated with the contact duration). Our contact diary data revealed that individuals with many close-range contacts on average spend a similar amount of time per contact as those with few close-range contacts. The evidence suggests that the augmented risk associated with greater contact numbers are not offset by shorter durations. While these findings reinforce the critical role hubs may play in disease propagation, lack of data on contact duration for individuals with (much) higher contact volume than those we could observe prevent us from identifying the point above which a negative correlation between contact volume and average contact duration appears. While theoretical studies that make the probability of transmission inversely contingent on a node's degree do not univocally find that this negative correlation attenuates the importance of hubs (Olinky and Stone 2004), in order to settle this question we need better data on the relationship between transmission probability, contact duration, and contact frequency.

**Appendix**

## A1. Variations of close-range contact heterogeneity by gender, age, and profession

We documented high variability in the number of close-range contacts across COMES-F's respondents (see figure 1). Here we show that this variability persists within major demographic categories. Figure A1a's upper panel shows the distribution of per day self-reported close-range contacts by respondent's gender. Past analyses of COMES-F data found that women (mainly adult women) tend to have a higher average number of contacts than men (see Béraud et al. 2015: 6 and table 1). Figure A1a shows that, behind this average difference, there exists a large degree of variation within genders. For both men and women numbers of contacts (far) higher than the median occur frequently.

Figure A1a's bottom panel shows the distribution of (diary-based) close-range contacts per day by respondent's age. Age is the most recurrent variable used in epidemiological models for representing socially structured social interactions. Age assortativity (and dissortativity at home) is found to be one of the most robust empirical regularities in epidemiological social contact surveys, as also found in multivariate analyses of the COMES-F data (see Béraud et al. 2015: 7-8). This motivates the use of average contacts per (more or less disaggregate) age-groups in age-structured compartment models (see, for some recent examples, Di Domenico et al. 2020; Roux et al. 2020; Salje et al. 2020: 3-4; in a comparative perspective, see Walker et al. 2020). Net of main effects of age on the likelihood of having more social contacts (which is indeed found in these data, see Béraud et al. 2015: table 1), figure A1a's bottom panel again shows high variability within age-groups.

Focusing the analysis on adult respondents in employment, we find a similar pattern for broad occupational groups (see figure A1b). Among both diary-based contacts (top panel) and extra job-related contacts declared by (bottom panel), there is a great deal of within-group variation across individuals in the number of self-reported close-range contacts. High-contact individuals seem especially concentrated among high (e.g. elementary school teacher, teaching assistant) and low routine non-manual workers (e.g. bank teller, teller in public administration) and service class (e.g. university professor, politician, journalist or doctor).

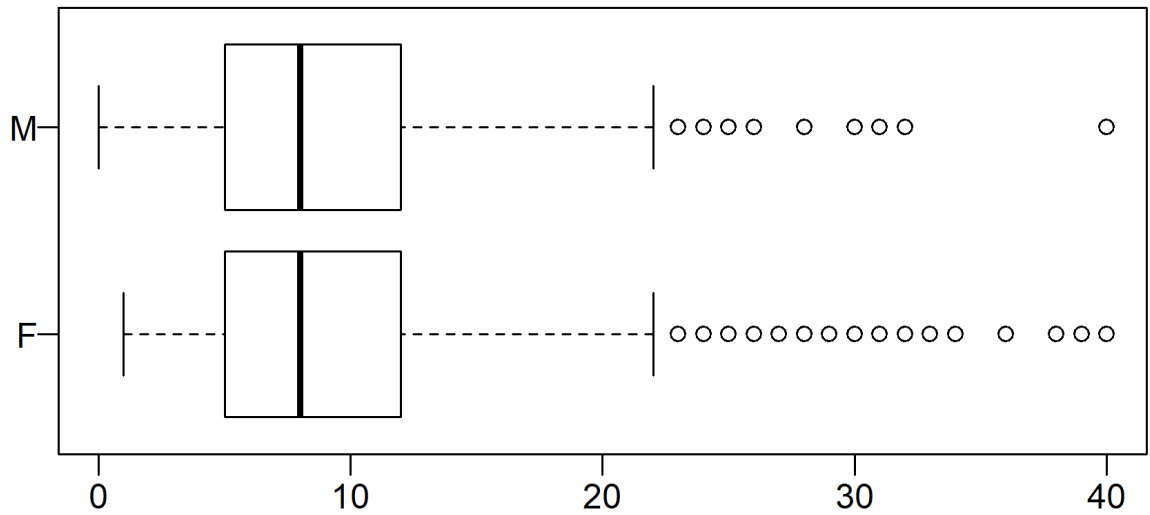

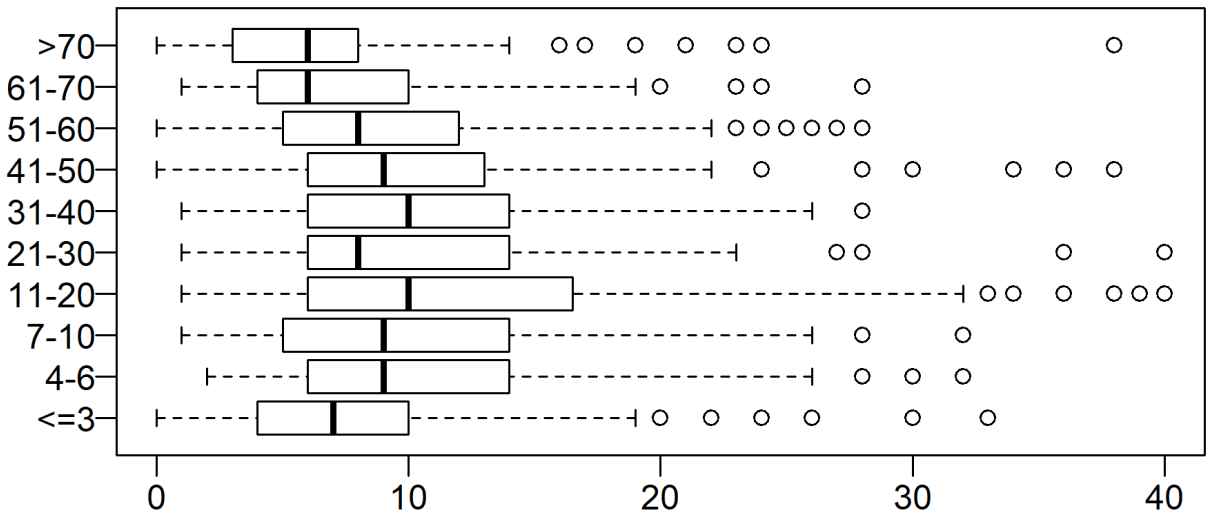

**Figure A1a.** Distribution of self-reported per day close-range contacts (x-axis) by gender (top panel; F [1136], M [897]) and age groups (bottom panel;<=3 [240], 4-6 [169], 7-10 [196], 11-20 [276], 21-30 [155], 31-40 [109], 41-50 [135], 51-60 [195], 61-70 [357], >70 [201]) (y-axis).

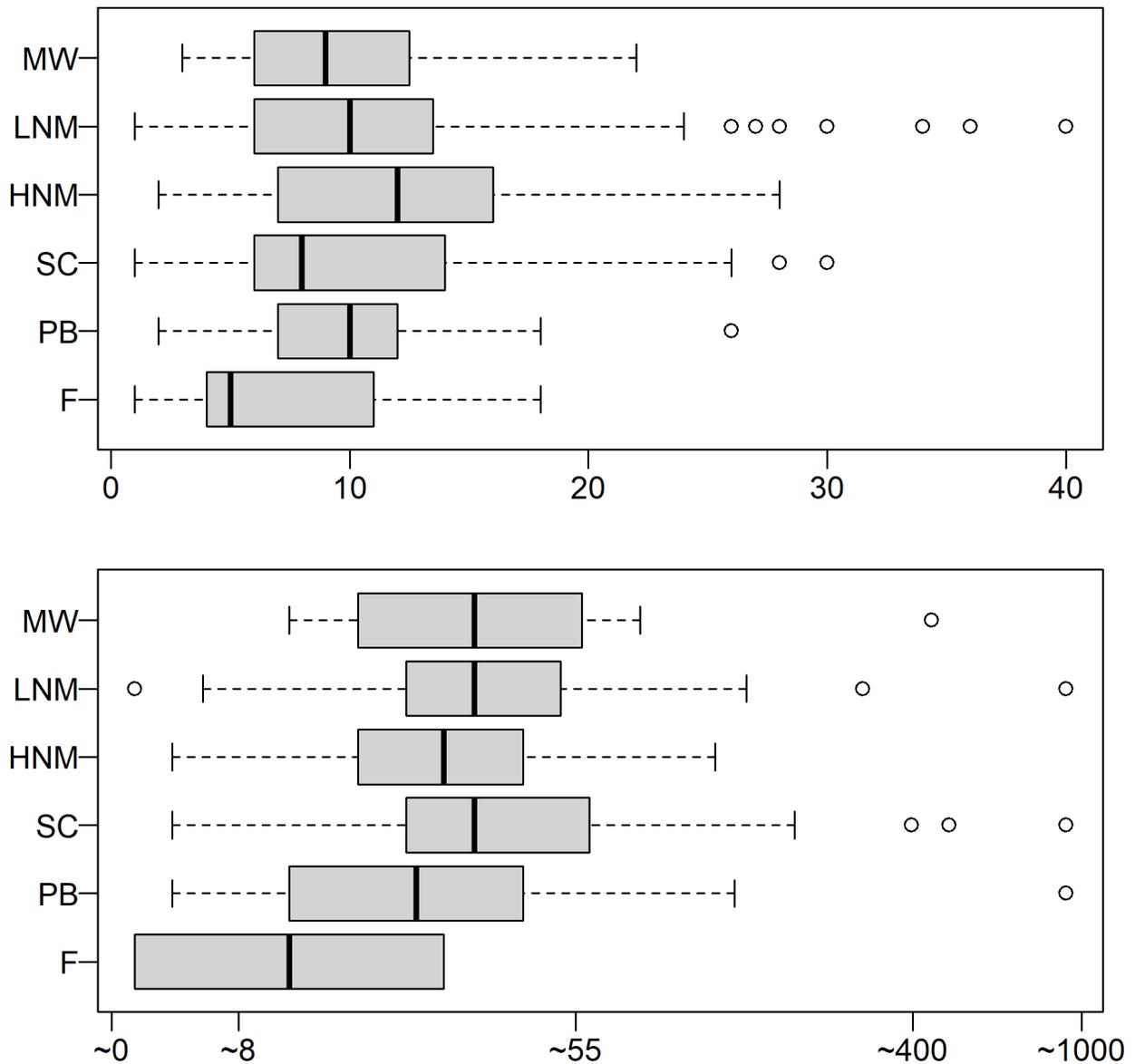

**Figure A1b.** *Top*: Distributions of daily close-range contacts among employed respondents (n=436) by occupational category (y-axis); *Bottom* (logarithmic scale with labels corresponding to values on the original linear scale): Distribution of daily job-related contacts among respondents regarding their occupation as especially exposed to close-range contact (*n*=257) by occupation category (y-axis). Occupational categories (*n* in parenthesis, top panel first): F=farmers ($n_{TOP}$=14, $n_{BOTTOM}$=2), PB=petty bourgeoisie (craftsmen and shopkeepers) and entrepreneurs ($n_{TOP}$=26, $n_{BOTTOM}$=18), SC= Service class (managers, high-skilled administrators, intellectual, scientific and liberal professions) ($n_{TOP}$=123, $n_{BOTTOM}$=91), HNM=High routine non-manual worker ($n_{TOP}$=47, $n_{BOTTOM}$=41), LNM=Low routine non-manual workers ($n_{TOP}$=183, $n_{BOTTOM}$=93), MW=manual workers ($n_{TOP}$=43, $n_{BOTTOM}$=12).

While these data suggest that social contacts within occupations are much more dispersed than one could expect under a distribution symmetrically centered around the mean, COMES-F does not provide a detailed list of jobs. The professional categories in the COMES-F data are too coarse to evaluate the effectiveness of a method for targeting hubs on the basis of employment status. We therefore exploit a recent survey that has a somewhat less thorough measurement of close-range contact (Belot et al. 2020) but fine-grained professional categories.

The Belot et al. (2020) survey was conducted in the third week of April, 2020 in the midst of the Covid-19 epidemic in six countries: China, South Korea, Japan, Italy, the UK and four states in the US: California, Florida, New York, and Texas. The sample consists of roughly 1,000 individuals from each country for a total of 6,082 respondents. Data was collected using market research companies Lucid and dataSpring, using gender and income quota. With regard to close-range contact, instead of being asked to keep a two-day-long diary, respondents were asked: "On a typical working day (before the outbreak of Covid-19), with how many people would you have close social contact (at less than 1 meter distance) and how long would you interact with them? (indicate approximate numbers - leave blank if the answer is zero)". Respondents' professions were classified in terms of the O-Net classification used by the US Bureau of Labor Statistics.

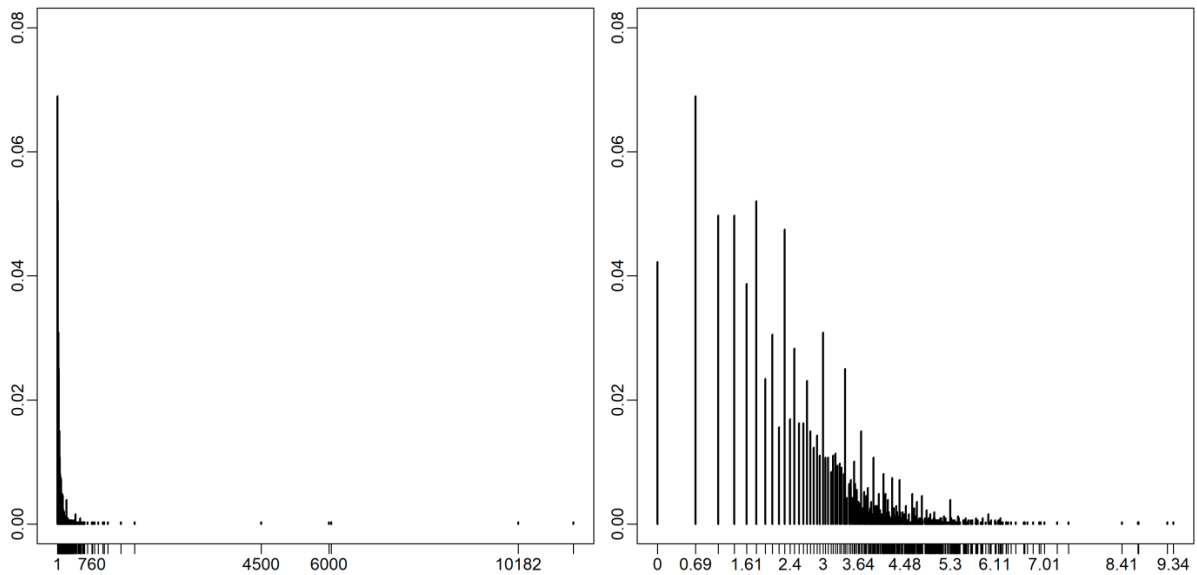

**Figure A1c.** Fraction of cases (y-axis) reporting a given number of close-range contacts (x-axis) in the Belot et al. (2020) data (n=4,103). Left: Linear scale. Right: Logarithmic scale.

The distribution of close contact frequency in the Belot et al. (2020) data is displayed in Figure A1c. The distribution is severely right-skewed, as we also observed for the French survey data. This provides confidence that the existence of hubs is not a measurement artifact but a robust feature of contact networks: Using different methods for measuring contact the same distributional characteristic is obtained.

Table A1 shows the mean and median number of close-range contacts by profession, in descending order of mean contact frequency, combining short- and long-duration contacts, excluding zero answers and professions with 15 or fewer cases. Table A1 has face validity, topped by professions that clearly involve close contact with many individuals -- elementary school teachers, cashiers -- and at the very bottom individuals who mostly work from home -- computer programmers. Some professions have an order of magnitude greater mean close-range contact than others. The spread is substantial especially when considering the ambiguity in the possible interpretation of the phrase "social contact" used in the questionnaire, translated into different languages, and the difficult task of estimating such numbers without use of a contact diary, which may produce noise that suppresses measured occupational differences.

Thus, given the systematic covariation that seems to exist between the fraction of high-contact individuals and specific occupation, when searching for hubs, targeting selected professions may be a reasonably effective strategy for finding hubs in contact networks.

| Profession | Mean # contacts | Median # contacts | N |
|---|---|---|---|
| Elementary School Teacher | 120 | 50 | 17 |
| Cashier | 76 | 40 | 20 |
| Order Clerks | 70 | 34 | 34 |
| Teacher Assistants | 67 | 40 | 18 |
| Retail Salespersons | 62 | 17 | 29 |
| Administrative Services Managers | 59 | 16 | 47 |
| Childcare Workers | 49 | 30 | 19 |
| Bill and Account Collectors | 45 | 19 | 21 |
| Sales Managers | 45 | 17 | 20 |
| Computer and Information Systems Managers | 34 | 14 | 33 |
| Financial Analysts | 34 | 19 | 25 |
| Customer Service Reps | 31 | 18 | 49 |
| Audio and Video Equipment Technicians | 31 | 19 | 27 |
| Construction Managers | 30 | 10 | 33 |
| Construction Laborers | 28 | 12 | 24 |
| Architectural Drafters | 25 | 11 | 16 |
| File Clerks | 24 | 12 | 38 |
| Civil Engineers | 23 | 16 | 53 |
| Data Entry Keyers | 22 | 16 | 17 |
| Credit Checkers | 22 | 14 | 16 |
| Ophthalmic Laboratory Technicians | 21 | 14 | 18 |
| Computer Network Support Specialists | 19 | 12 | 28 |
| Financial Managers, Branch or Department | 17 | 7 | 25 |
| Financial Examiners | 17 | 6 | 20 |
| Computer Programmers | 15 | 9 | 16 |

**Table A1.** Close contact by profession in Belot et al. data (2020)

## A2. Results for the Erdős–Rényi (ER) network and for the Degree-calibrated (DC) networks under various transmission probabilities

Figure 4 and table 3 show large differences in the effectiveness of interventions that do and do not target high-contact individuals for intervention. Here we explore how instrumental the skewness in the empirical distribution of close-range contact is for the effectiveness of hub targeting. We do so by recalculating the figure 4 and table 3 results for the Erdős–Rényi (ER) network with the same average degree as the empirical degree distribution (see table 1), respectively figure A2a and table A2.

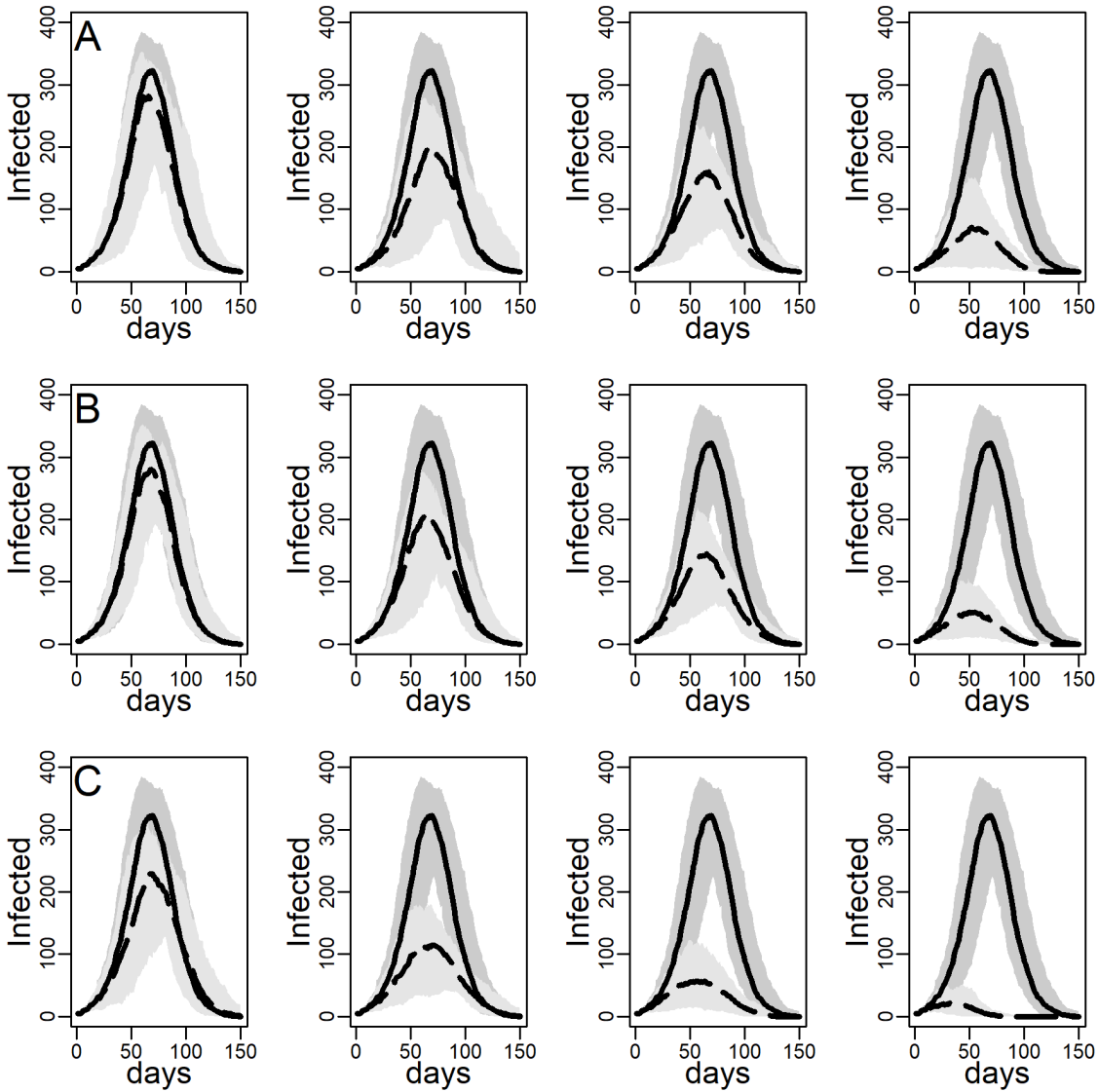

**Figure A2a.** Number of infected agents (y-axis) by days (x-axis) (median of 100 replications) under three different interventions (rows) targeting 1, 3, 5, or 10 agents per day (columns). A – NO-TARGET intervention; B –

CONTACT-TARGET intervention. C – HUB-TARGET intervention. Lower and upper bounds of the shaded areas correspond to the 5[th] percentiles and 95[th] percentiles of the 100 replications. Solid line: *Erdős–Rényi* (ER) network; dashed line: interventions. Dyadic transmission probability *r*=0.05. n = 2,029 agents.

A comparison of panels A between figures A2 and 4 shows that NO-TARGET interventions are less effective in DC networks with high degree skew than in ER networks with low degree variance. This suggests that models that do not account for empirical network structure may overestimate the expected impact of interventions. Comparing panels B and C across figures we find that HUB-TARGET and CONTACT-TARGET interventions are much more effective in the DC network than in ER networks, where the to-be-immunized agents have lower network degree.

| *Erdős–Rényi* (ER) network | | | | | | | no intervention: H = 322 [208.65; 376.35] T = 68 | |
|---|---|---|---|---|---|---|---|---|
| | | *b* = 1 | | *b* = 3 | | *b* = 5 | | *b* = 10 |
| | peak height | time | peak height | time | peak height | time | peak height | time |
| NO-TARGET | 281.5 [141.6; 340.25] | 66 | 200 [58.75; 264.40] | 69 | 160.5 [60; 214.05] | 67 | 71.5 [3.95; 150.15] | 52 |
| CONTACT-TARGET | 281 [178.55; 338.15] | 68 | 208 [73.9; 271.05] | 65 | 145 [51.95; 204] | 65 | 51.5 [10.85; 98.25] | 50 |
| HUB-TARGET | 229 [95.85 ; 302.20] | 66 | 114.5 [33.9; 175.35] | 70 | 56.5 [14.95; 118.009] | 56 | 22 [3; 43.35] | 38 |

**Table A2**. Peak height (maximum # concurrently infected agents) and time (in days) under three interventions (rows) and four budgets (column) on the *Erdős–Rényi* (ER) network. Dyadic transmission probability *r*=0.05. Shown are median, 5% and 95% percentiles across 100 iterations.

We next check the stability of the results shown in Figure 4 and table 3 about the effectiveness of interventions that do and do not target high-contact individuals for intervention on the DC network when we decrease/increase the infectioness of the disease. To this aim, below, we recalculate figure 4-6 under low and high dyadic transmission probabilities *r*, in both cases over the entire range of clustering levels we built in our DC contact networks. Results are qualitatively unchanged. The relative differences across targeting methods are attenuated for small-size epidemics triggered by low dyadic transmission probabilities (see figs. A2b-d) whereas they are enhanced for larger epidemics associated with high dyadic transmission probabilities (figs. A2e-f). However, the effectiveness of the contact- and hub-targeting methods in mitigating the epidemic relatively to the random targeting method is still observed across all combinations of transmission probability and clustering levels.

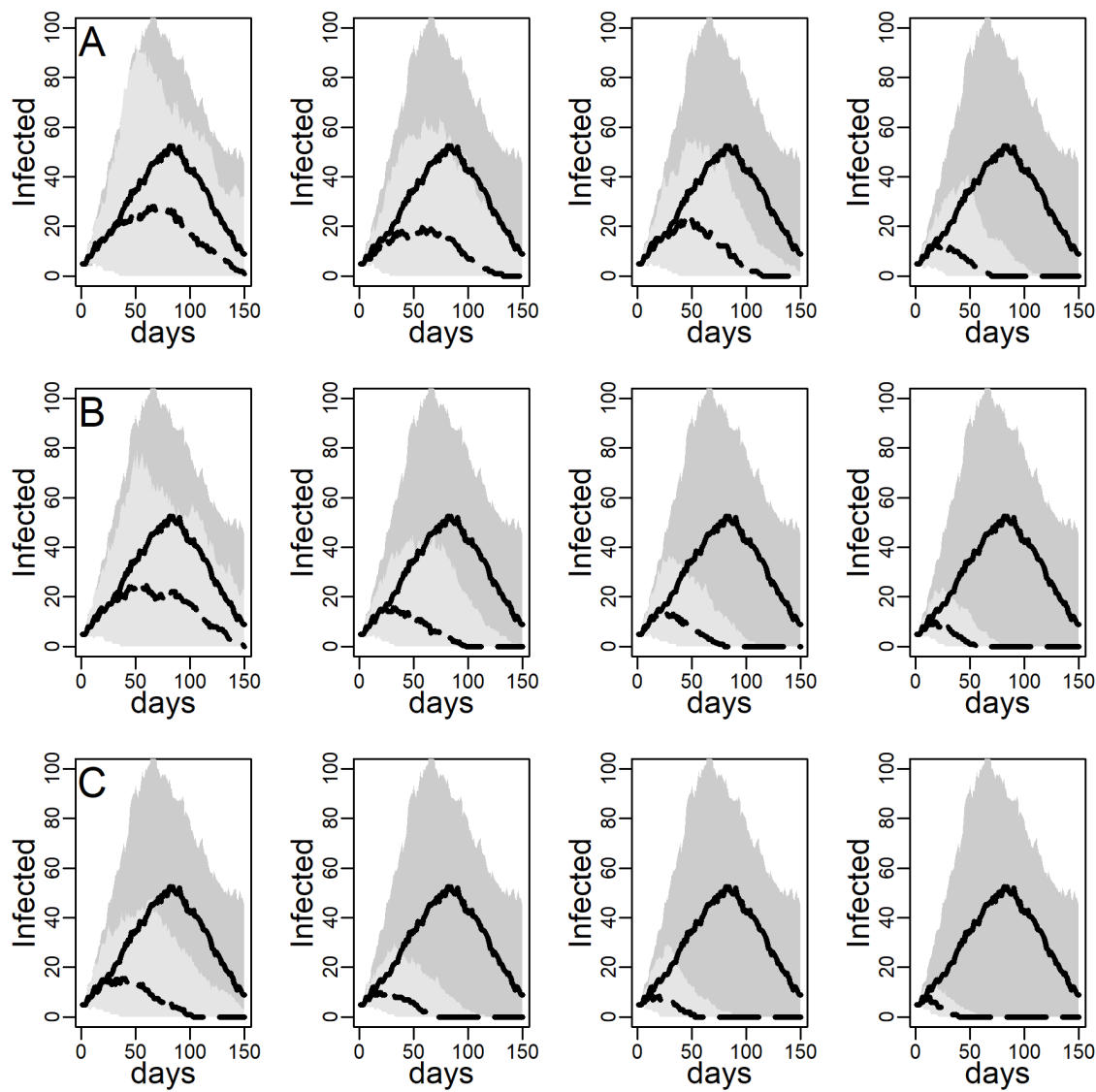

**Figure A2b.** Number of infected agents (y-axis) by days (x-axis) (median of 100 replications) under three different interventions (rows) targeting 1, 3, 5, or 10 agents per day (columns). A – NO-TARGET intervention; B – CONTACT-TARGET intervention. C – HUB-TARGET intervention. Lower and upper bounds of the shaded areas correspond to the 5[th] percentiles and 95[th] percentiles of the 100 replications. Solid line: *Degree-Calibrated* (DC) network; dashed line: interventions. Dyadic transmission probability *r*=0.03 & Local clustering (Cc=0.01). *n* = 2,029 agents.

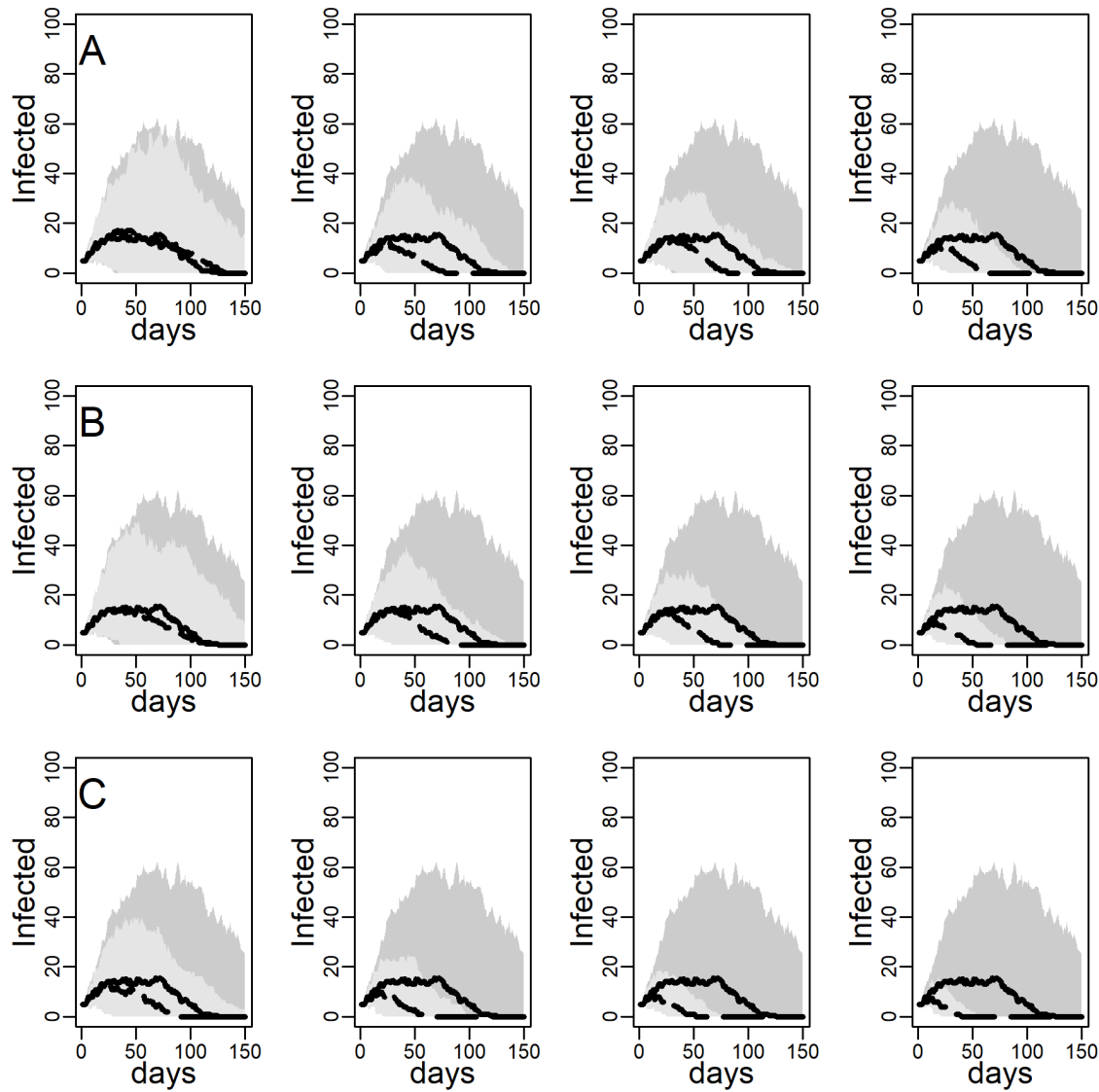

**Figure A2c.** Number of infected agents (y-axis) by days (x-axis) (median of 100 replications) under three different interventions (rows) targeting 1, 3, 5, or 10 agents per day (columns). A – NO-TARGET intervention; B – CONTACT-TARGET intervention. C – HUB-TARGET intervention. Lower and upper bounds of the shaded areas correspond to the 5th percentiles and 95th percentiles of the 100 replications. Solid line: *Degree-Calibrated* (DC) network; dashed line: interventions. Dyadic transmission probability $r$=0.03 & Local clustering (Cc=0.43). $n$ = 2,029 agents.

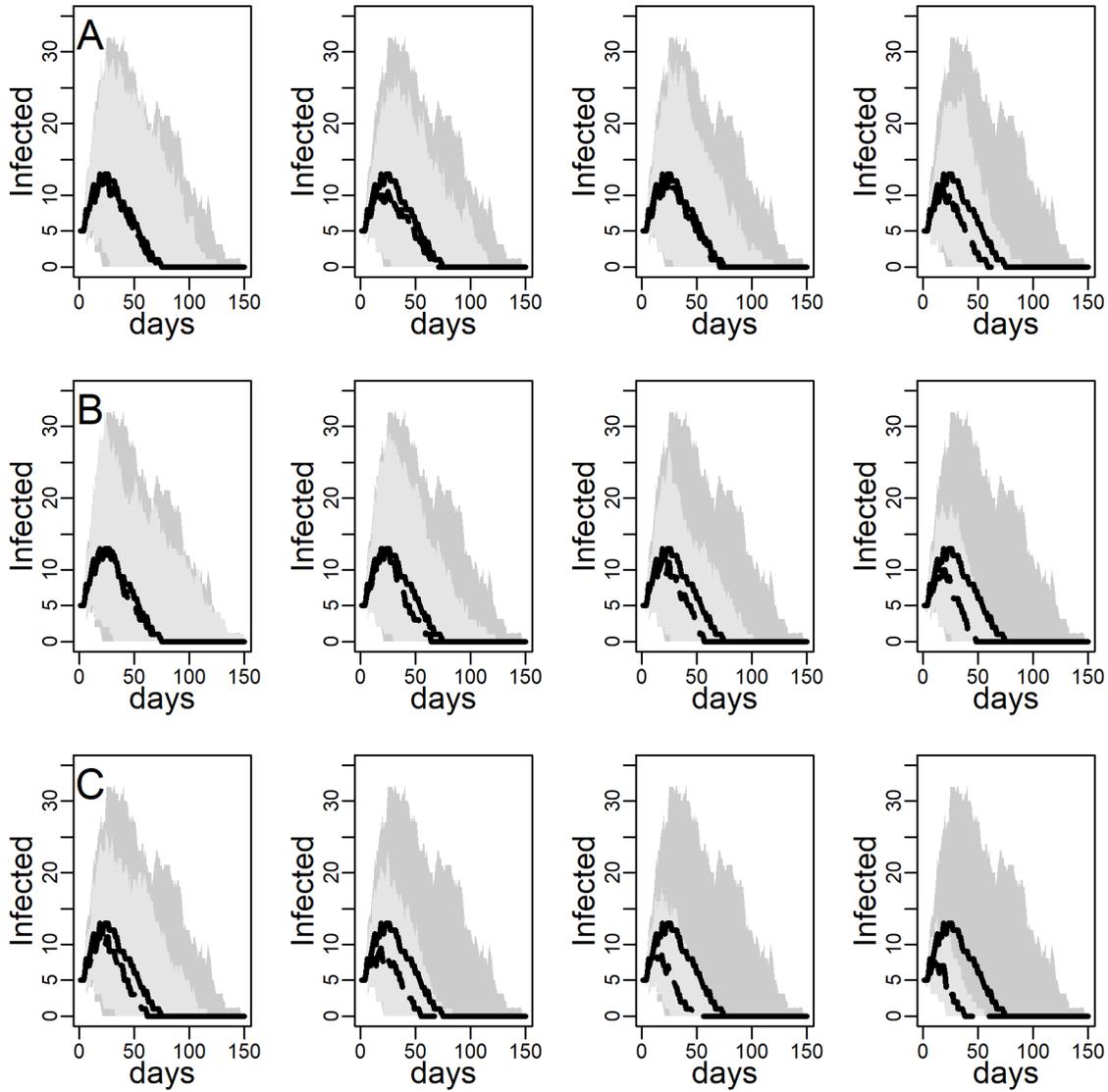

**Figure A2d.** Number of infected agents (y-axis) by days (x-axis) (median of 100 replications) under three different interventions (rows) targeting 1, 3, 5, or 10 agents per day (columns). A – NO-TARGET intervention; B – CONTACT-TARGET intervention. C – HUB-TARGET intervention. Lower and upper bounds of the shaded areas correspond to the 5[th] percentiles and 95[th] percentiles of the 100 replications. Solid line: *Degree-Calibrated* (DC) network; dashed line: interventions. Dyadic transmission probability *r*=0.03 & Local clustering (Cc=0.57). *n* = 2,029 agents.

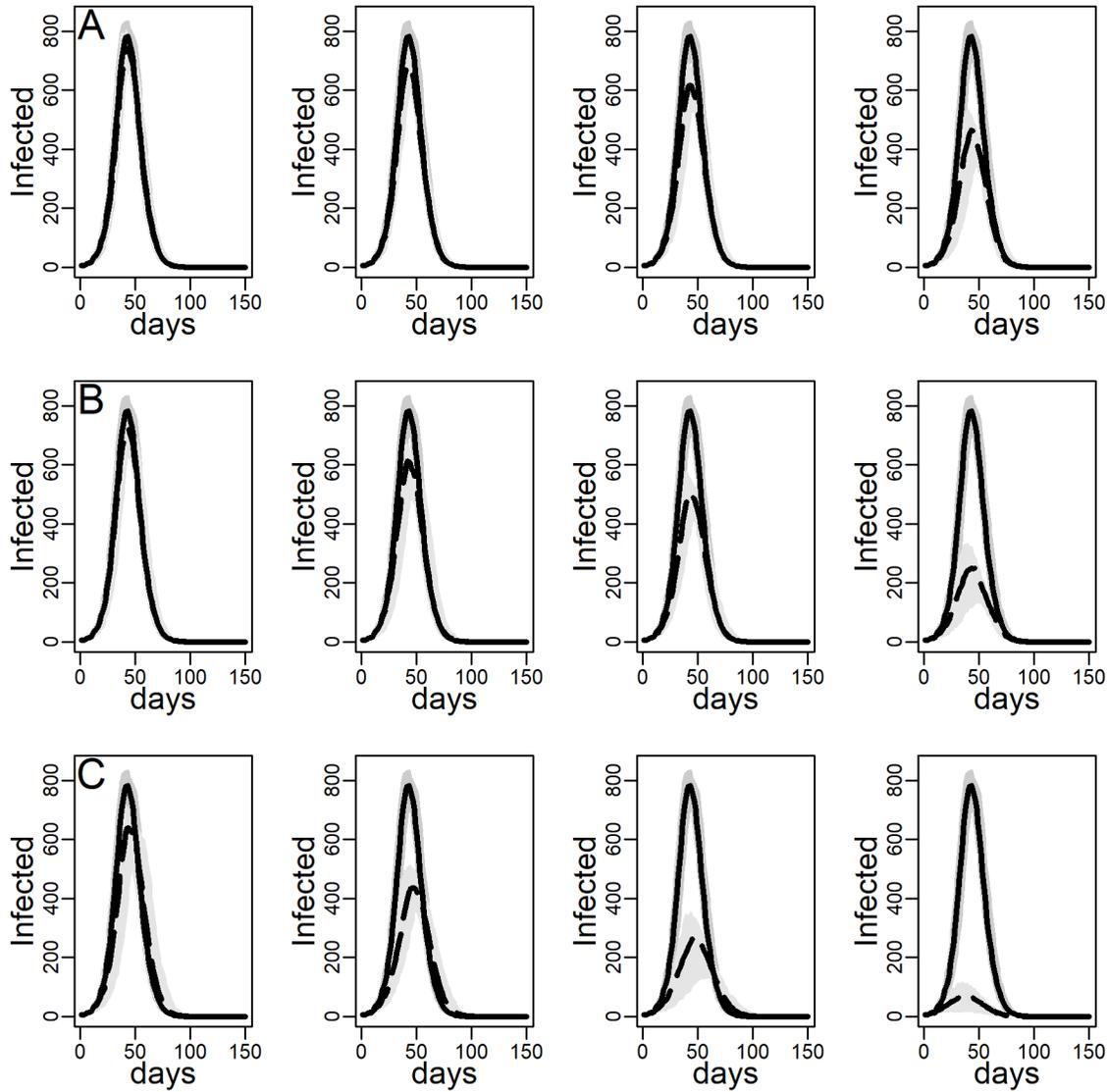

**Figure A2e.** Number of infected agents (y-axis) by days (x-axis) (median of 100 replications) under three different interventions (rows) targeting 1, 3, 5, or 10 agents per day (columns). A – NO-TARGET intervention; B – CONTACT-TARGET intervention. C – HUB-TARGET intervention. Lower and upper bounds of the shaded areas correspond to the 5[th] percentiles and 95[th] percentiles of the 100 replications. Solid line: *Degree-Calibrated* (DC) network; dashed line: interventions. Dyadic transmission probability *r*=0.07 & Local clustering (Cc=0.01). *n* = 2,029 agents.

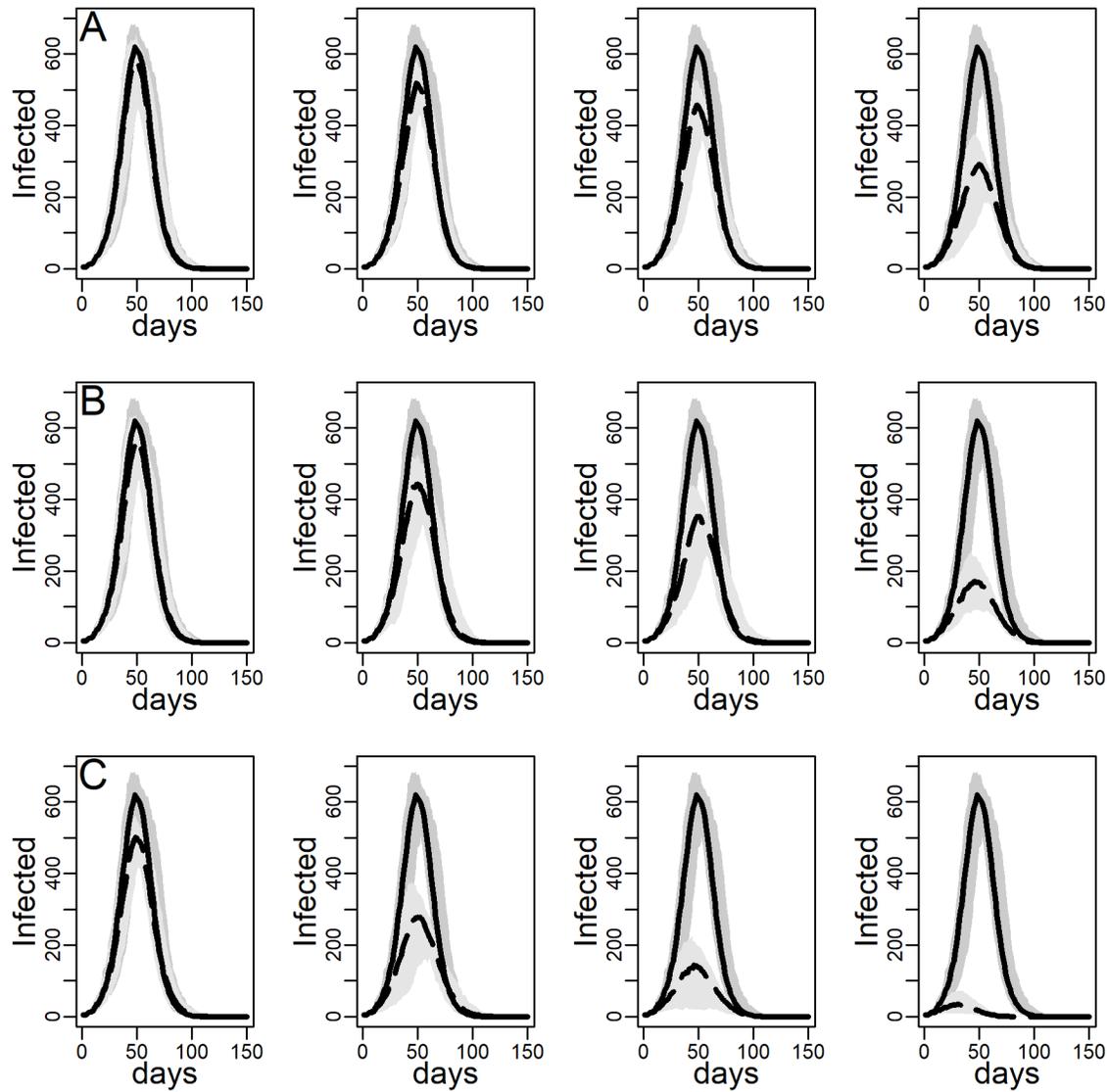

**Figure A2f.** Number of infected agents (y-axis) by days (x-axis) (median of 100 replications) under three different interventions (rows) targeting 1, 3, 5, or 10 agents per day (columns). A – NO-TARGET intervention; B – CONTACT-TARGET intervention. C – HUB-TARGET intervention. Lower and upper bounds of the shaded areas correspond to the 5[th] percentiles and 95[th] percentiles of the 100 replications. Solid line: *Degree-Calibrated* (DC) network; dashed line: interventions. Dyadic transmission probability *r*=0.07 & Local clustering (Cc=0.43). *n* = 2,029 agents.

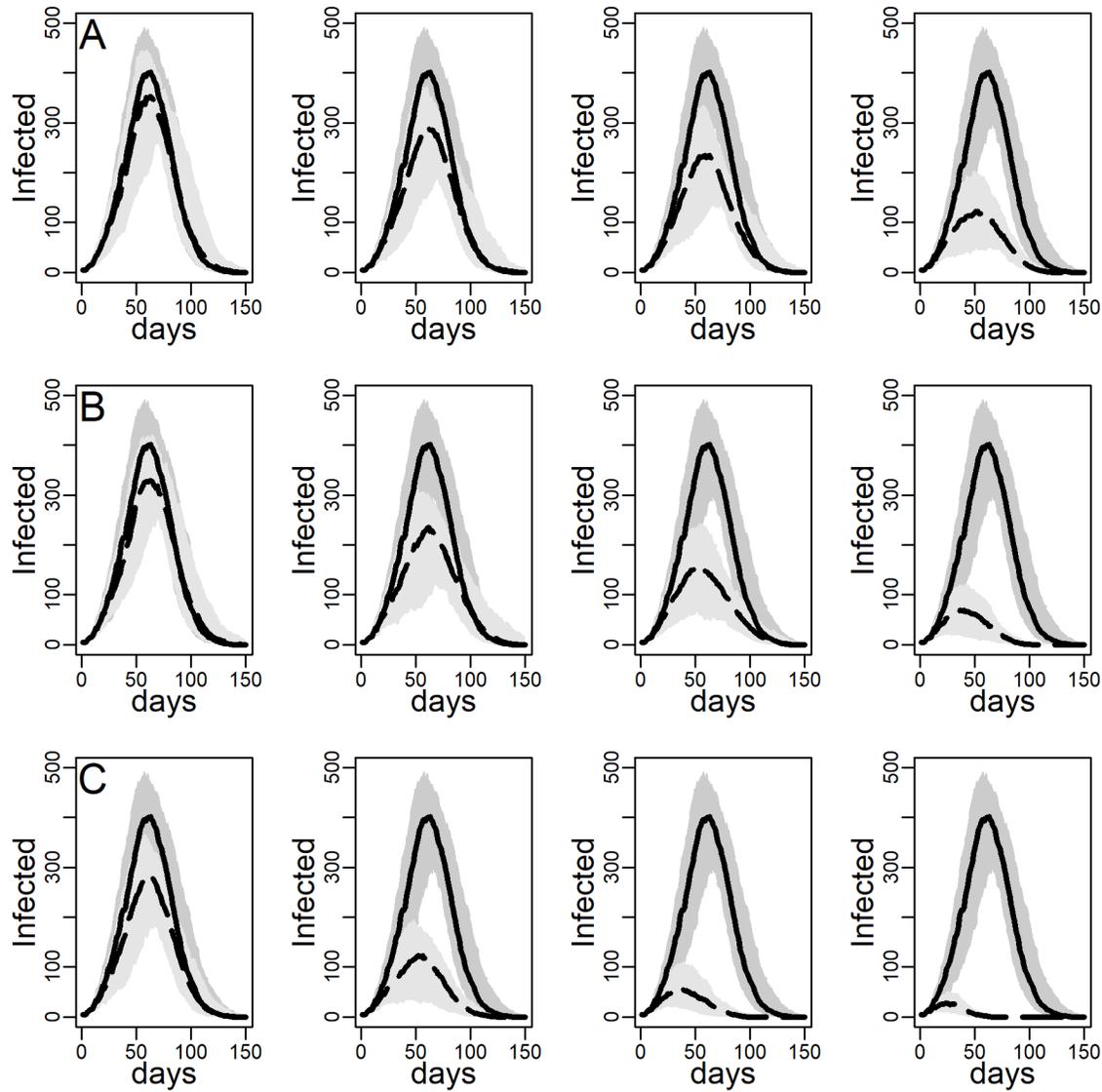

**Figure A2g.** Number of infected agents (y-axis) by days (x-axis) (median of 100 replications) under three different interventions (rows) targeting 1, 3, 5, or 10 agents per day (columns). A – NO-TARGET intervention; B – CONTACT-TARGET intervention. C – HUB-TARGET intervention. Lower and upper bounds of the shaded areas correspond to the 5[th] percentiles and 95[th] percentiles of the 100 replications. Solid line: *Degree-Calibrated* (DC) network; dashed line: interventions. Dyadic transmission probability *r*=0.07 & Local clustering (Cc=0.57). *n* = 2,029 agents.

## A3. Results for the Degree-Calibrated (DC) network using additional professional contact data

Here we report on results obtained by calibrating the synthetic network through a measure of daily close contact for each respondent that combines diary-based contacts (figure 1's left plot) and job-related extra contacts (figure 2's left plot). In particular, for respondents in employment who self-reported job-related extra contacts (n=259), these contacts are summed up to the number of contacts recorded through the diary. However, we limited the portion of job-related extra contacts to be added in such a way that the total number of contacts is never higher than 134. Existing contact survey data suggests indeed that the monotonic increase in the relationship between the number of close-range contacts and the total amount of time of these contacts that we described in figure 2 (left plot) starts to become decreasing above 100 contacts (see Danon et al. 2012: fig. S3c; Danon et al. 2013: fig. 2). This suggests that the relationship between the total number of contacts and the average contact length may start to become negative above this threshold. We have chosen the specific value of 134 to be consistent with previous studies of COMES-F data, where supplementary professional contacts were censored at 134 (see Béraud et al. 2015: 5, 9). As a by-product, this choice prevents a few nodes (like the two with 500 or the three with 999 job-related contacts) from having contacts with a substantial fraction of our simulated population of ~2k agents, thus reducing the risk of artificially overestimating the impact of hubs.

Table A3 shows network statistics computed over 100 realizations of the DC and ER networks. Compared to the networks including only diary-based contacts (see table 1), apart from a higher average degree, it is noteworthy the larger standard-deviation (approximately 19 *versus* approximately 7) of the DC network, which reflects the larger portion of high-contact respondents that are now in the network. If one considers the right-tail of the degree distribution of this network (which again is well approximated by a power law with a scale parameter 2.54 for respondents with close-range contact above 17), this tail now contains 445 nodes (compared to 175 nodes for the power-law-like right-tail of the network including only diary-based contacts). Compared to the diary-based contact network, such a ticker right-tail translates into a lower average path length (even when built-in clustering increases), which well shows that the larger, and more numerous, the hubs the stronger their capacity to create bridges across otherwise distant parts of the network (on the connection between hubs and small-world behaviors, see Albert et al. 2000).

|  | **Average degree** | **Median degree** | **Stdev degree** | **Clustering coef** | **Deg-clust corr** | **Av path length** | **Diameter** |
|---|---|---|---|---|---|---|---|
| *Empirical-degree* (ED) networks | | | | | | | |
| *p*=0 | 14.87 (0.00) | 9.00 (0.00) | 19.58 (0.00) | 0.04 (0.00) | -0.15 (0.01) | 2.83 (0.00) | 4.8 (0.40) |
| *p*=0.5 | 14.72 (0.04) | 9.00 (0.00) | 19.19 (0.04) | 0.42 (0.01) | -0.50 (0.00) | 3.29 (0.02) | 5.01 (0.01) |
| *p*=1 | 14.77 (0.04) | 9.00 (0.00) | 19.38 (0.12) | 0.50 (0.01) | -0.44 (0.01) | 3.60 (0.08) | 6 (0.62) |
| *Erdős-Rényi* (ER) network | | | | | | | |
| ER | 14.86 (0.14) | 14.89 (0.31) | 3.86 (0.14) | 0.01 (0.01) | -0.00 (0.03) | 3.09 (0.01) | 5 (0.00) |

**Table A3.** Topological features of the simulated contact networks (as a function of the local tie probability *p*, for the *Degree-Calibrated* (DC) network). Mean values across 100 network realizations (standard deviation in parentheses). Clustering coef=clustering coefficient; Deg-clust corr=Pearson correlation coefficient between nodes' degree and their clustering coefficient; Av path length=Average of the shortest path lengths; Diameter=Maximum of the shortest path lengths.

We re-ran all the analyses on the extended DC networks and the ER network with the same average degree. Results are reported below. Figure A3a reproduces figure 3 in the main text; figures A3b-d (dyadic transmission probability r=0.05), A3e-g (dyadic transmission probability r=0.03), and A3h-l (dyadic transmission probability r=0.07) respectively reproduces figures 4-6 in the main text and figures A2b-d and A2e-g in appendix A2. We do not comment in detail on these figures because results are in line with our main analysis. As to the effects of degree skewness and clustering on epidemic's size and paste where no intervention is in place, (fig. A3a), we find indeed the same patterns we found on the network including only diary-based contact. And, as to the effectiveness of the contact- and hub-targeting methods in mitigating the epidemic relatively to the random targeting method, the relative gradient between these strategies is still observed across all combinations of transmission probability (r) and clustering levels. The superiority of targeting hubs only appears more clearly because of the larger size, and fraction, of high-contact nodes.

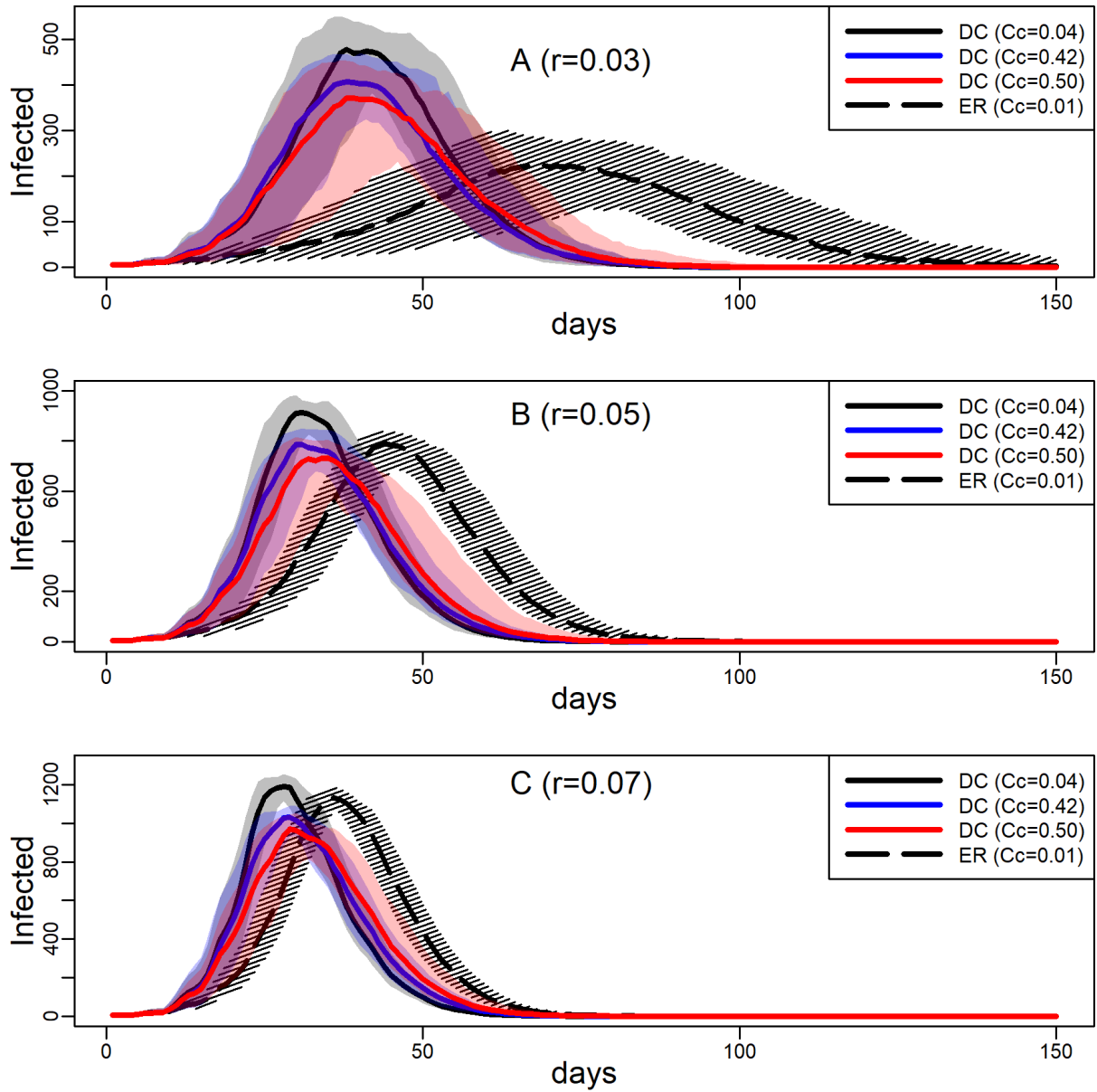

Figure A3a. Number of infected agents (y-axis) by days (x-axis) (median of 100 replications) as a function of increasing values of the dyadic transmission probability *r* and clustering (see Legend). Lower and upper bounds of the shaded areas correspond to the 5th percentiles and 95th percentiles of the 100 replications. n = 2,029 agents. Solid line: *Degree-Calibrated* (DC) networks; dashed line: *Erdős–Rényi* (ER) network with the same average degree.

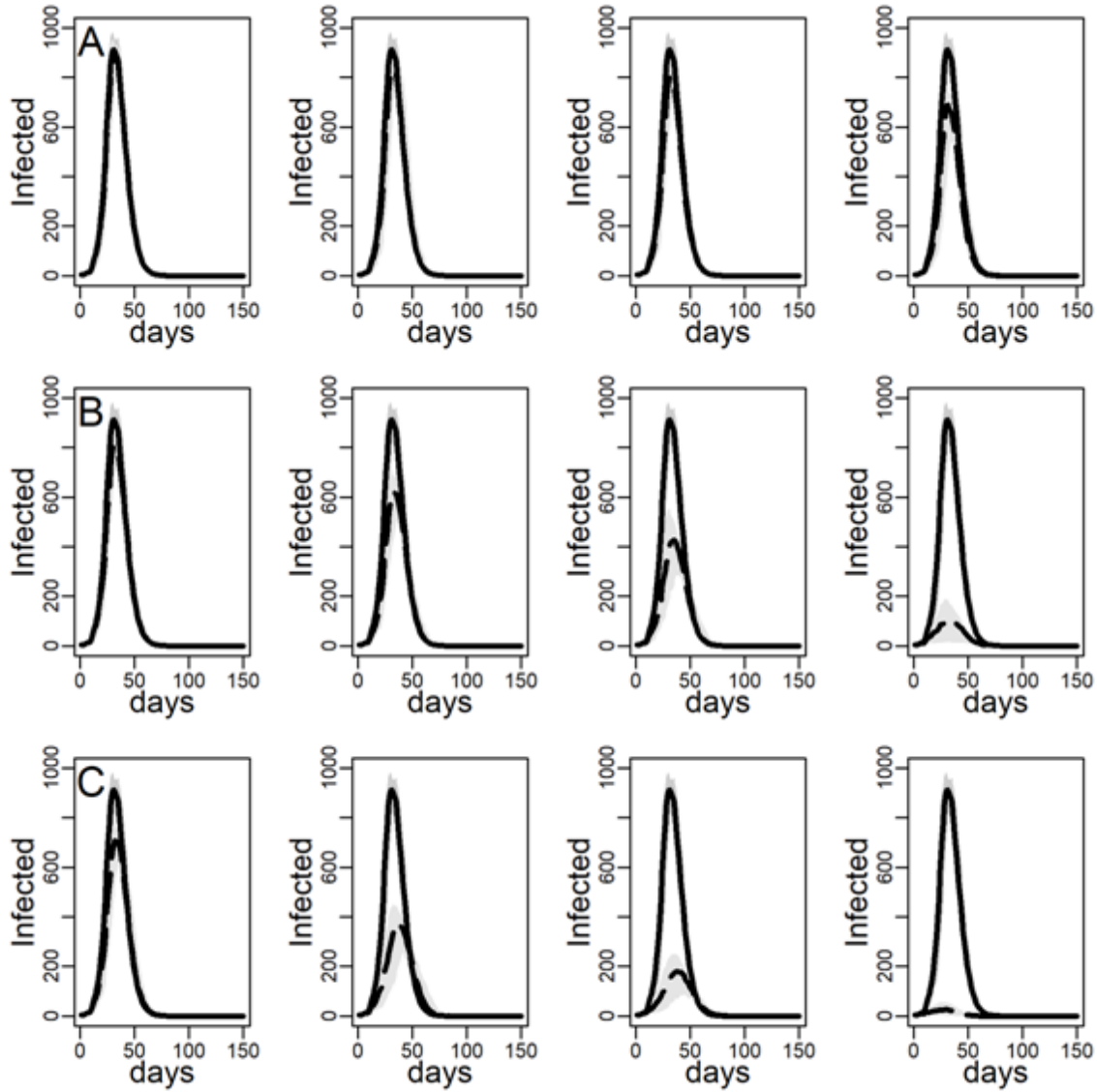

**Figure A3b.** Number of infected agents (y-axis) by days (x-axis) (median of 100 replications) under three different interventions (rows) targeting 1, 3, 5, or 10 agents per day (columns). A – NO-TARGET intervention; B – CONTACT-TARGET intervention. C – HUB-TARGET intervention. Lower and upper bounds of the shaded areas correspond to the 5[th] percentiles and 95[th] percentiles of the 100 replications. Solid line: *Degree-Calibrated* (DC) network; dashed line: interventions. Dyadic transmission probability (r=0.05) & Local clustering (Cc=0.04). *n* = 2,029 agents.

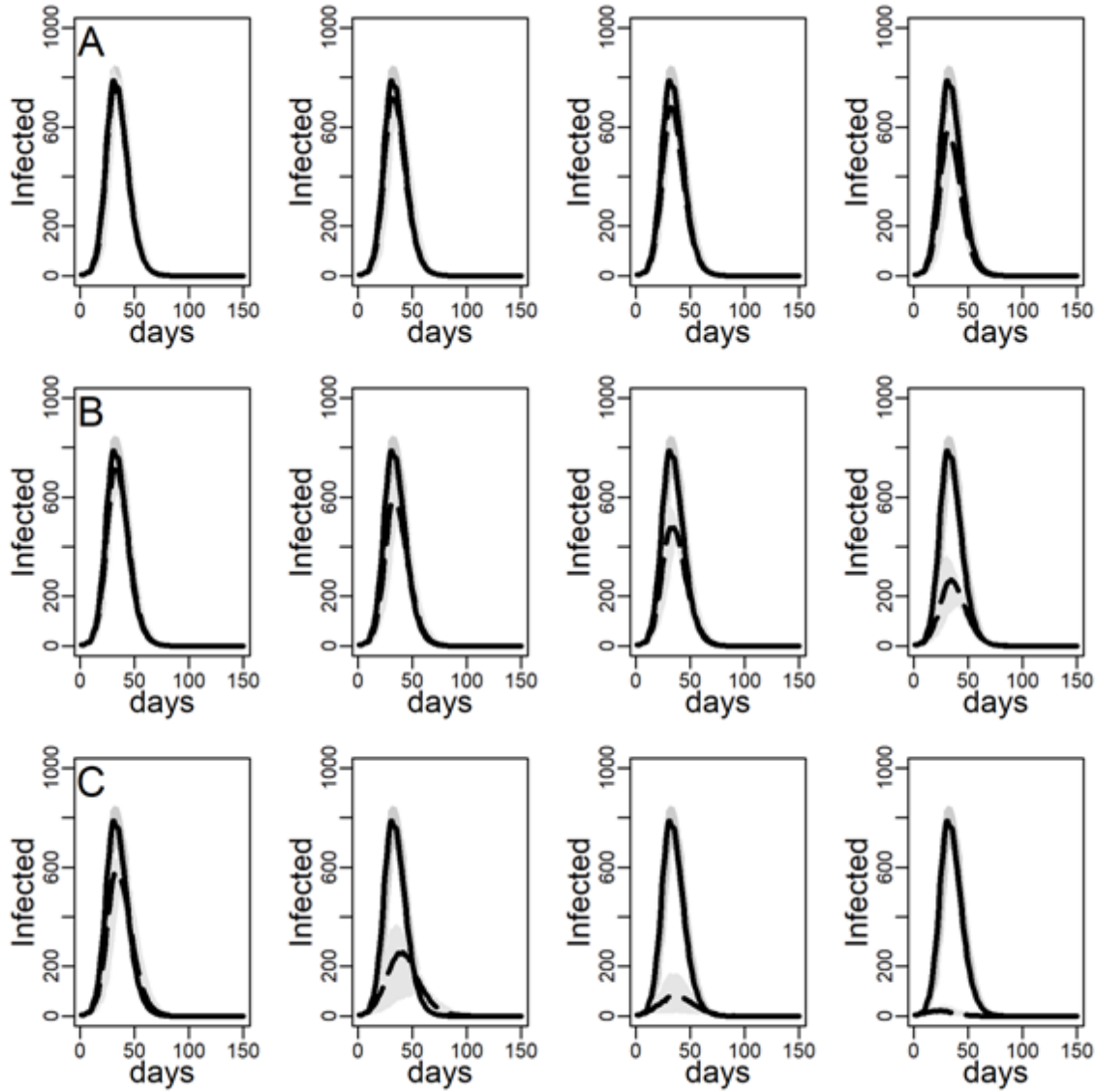

**Figure A3c.** Number of infected agents (y-axis) by days (x-axis) (median of 100 replications) under three different interventions (rows) targeting 1, 3, 5, or 10 agents per day (columns). A – NO-TARGET intervention; B – CONTACT-TARGET intervention. C – HUB-TARGET intervention. Lower and upper bounds of the shaded areas correspond to the 5[th] percentiles and 95[th] percentiles of the 100 replications. Solid line: *Degree-Calibrated* (DC) network; dashed line: interventions. Dyadic transmission probability (r=0.05) & Local clustering (Cc=0.42). *n* = 2,029 agents.

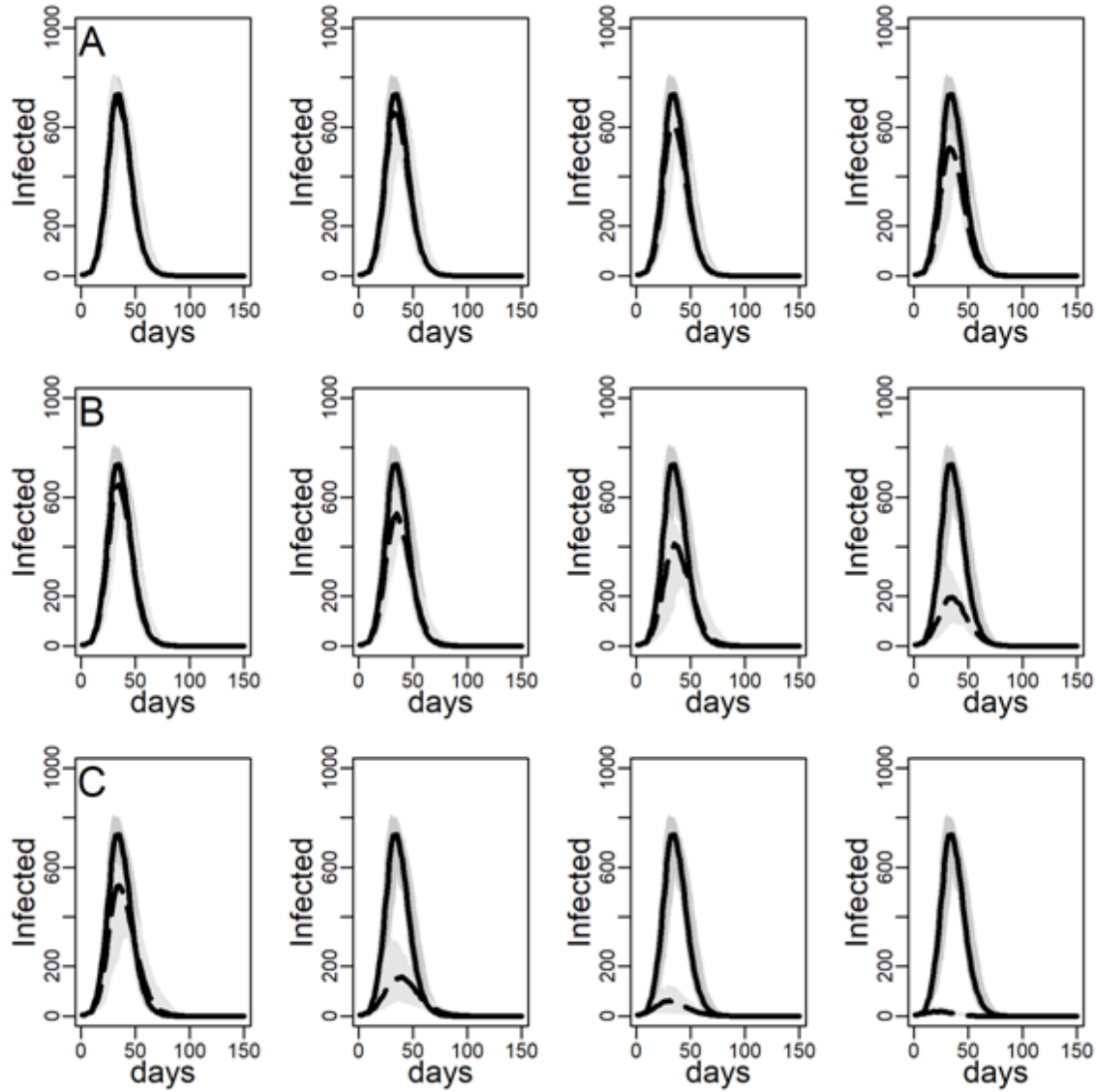

**Figure A3d.** Number of infected agents (y-axis) by days (x-axis) (median of 100 replications) under three different interventions (rows) targeting 1, 3, 5, or 10 agents per day (columns). A – NO-TARGET intervention; B – CONTACT-TARGET intervention. C – HUB-TARGET intervention. Lower and upper bounds of the shaded areas correspond to the 5[th] percentiles and 95[th] percentiles of the 100 replications. Solid line: *Degree-Calibrated* (DC) network; dashed line: interventions. Dyadic transmission probability (r=0.05) & Local clustering (Cc=0.50). *n* = 2,029 agents.

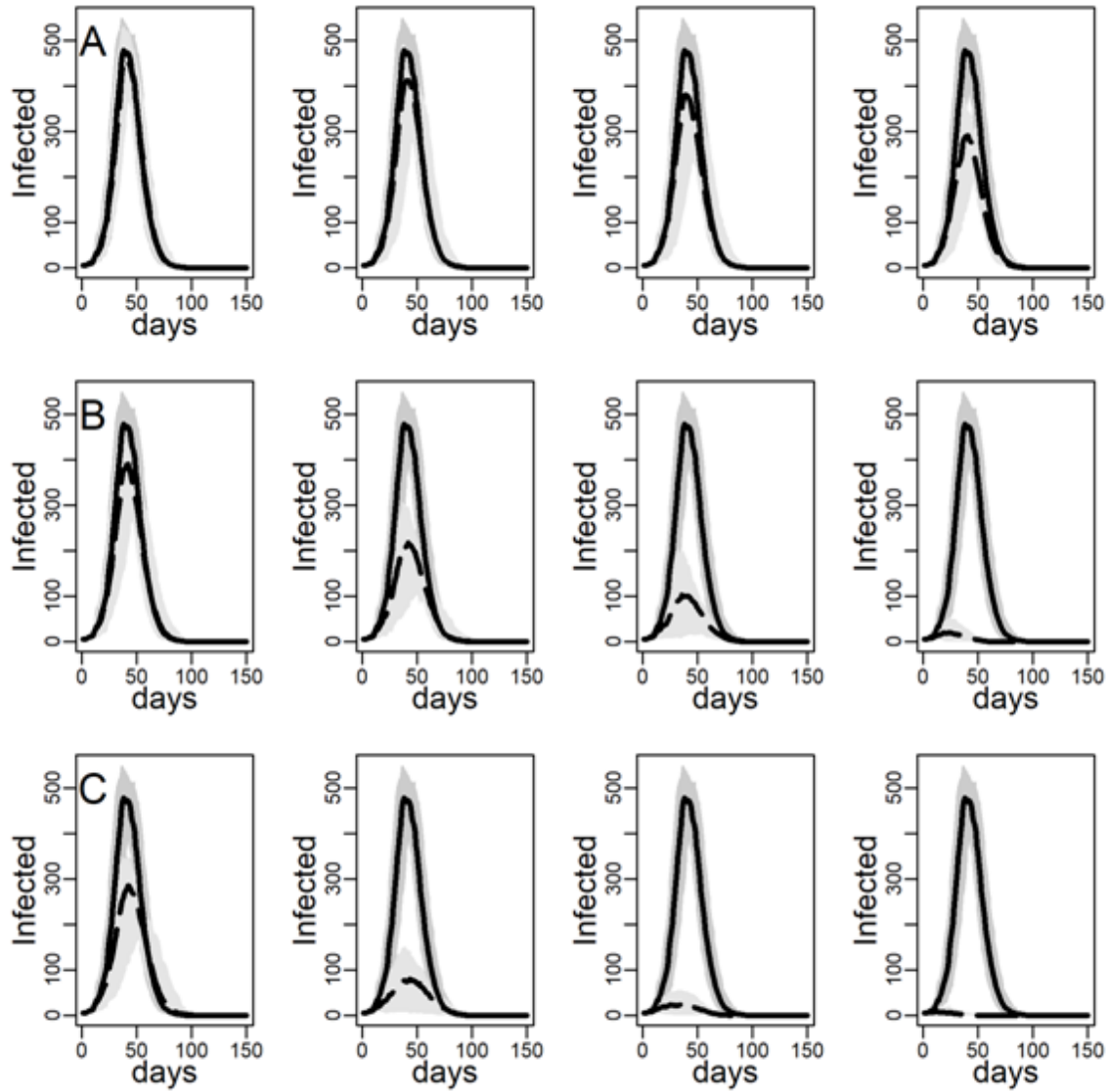

**Figure A3e.** Number of infected agents (y-axis) by days (x-axis) (median of 100 replications) under three different interventions (rows) targeting 1, 3, 5, or 10 agents per day (columns). A – NO-TARGET intervention; B – CONTACT-TARGET intervention. C – HUB-TARGET intervention. Lower and upper bounds of the shaded areas correspond to the 5[th] percentiles and 95[th] percentiles of the 100 replications. Solid line: *Degree-Calibrated* (DC) network; dashed line: interventions. Dyadic transmission probability (r=0.03) & Local clustering (Cc=0.04). *n* = 2,029 agents.

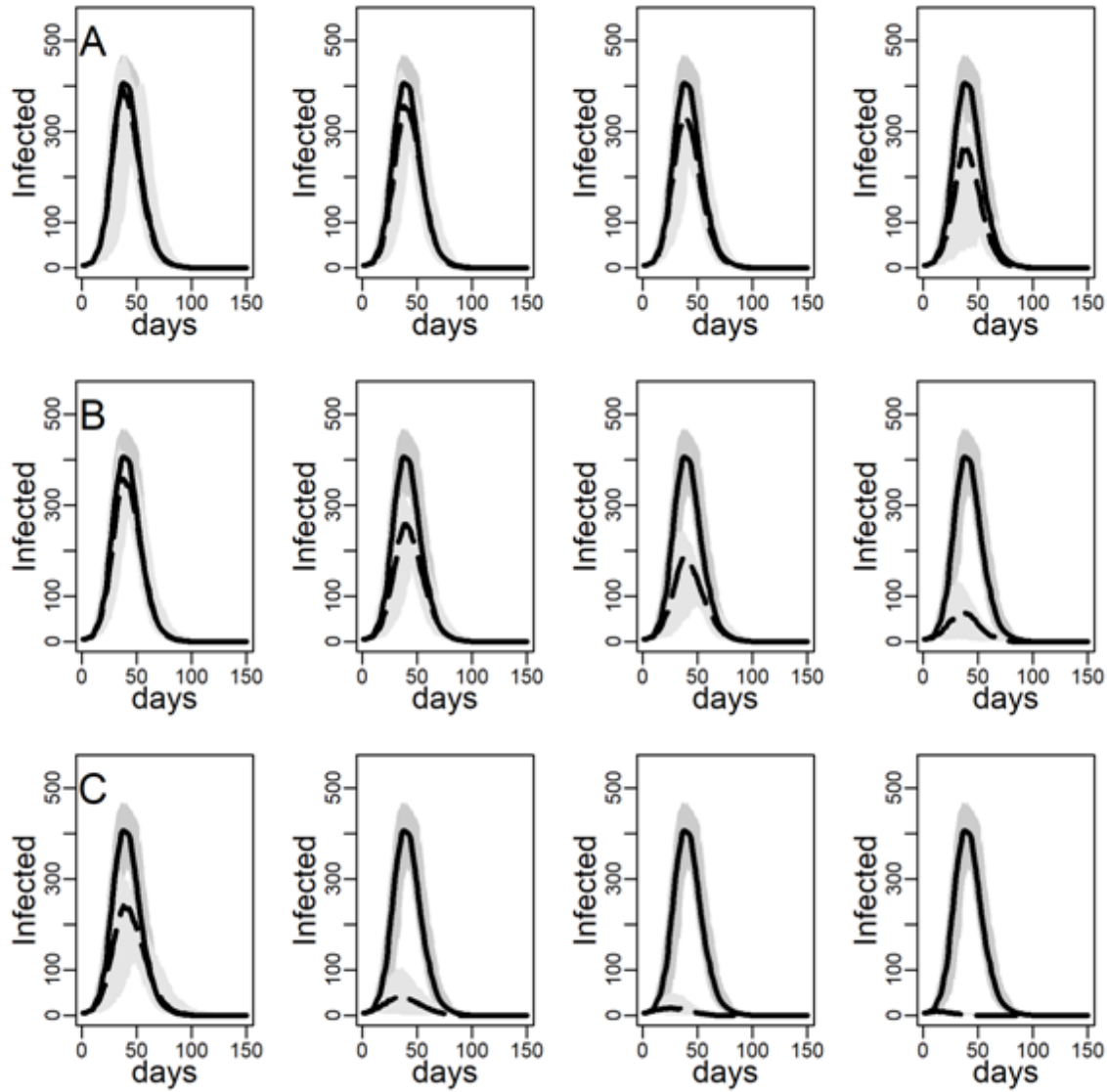

**Figure A3f.** Number of infected agents (y-axis) by days (x-axis) (median of 100 replications) under three different interventions (rows) targeting 1, 3, 5, or 10 agents per day (columns). A – NO-TARGET intervention; B – CONTACT-TARGET intervention. C – HUB-TARGET intervention. Lower and upper bounds of the shaded areas correspond to the 5th percentiles and 95th percentiles of the 100 replications. Solid line: *Degree-Calibrated* (DC) network; dashed line: interventions. Dyadic transmission probability (r=0.03) & Local clustering (Cc=0.42). *n* = 2,029 agents.

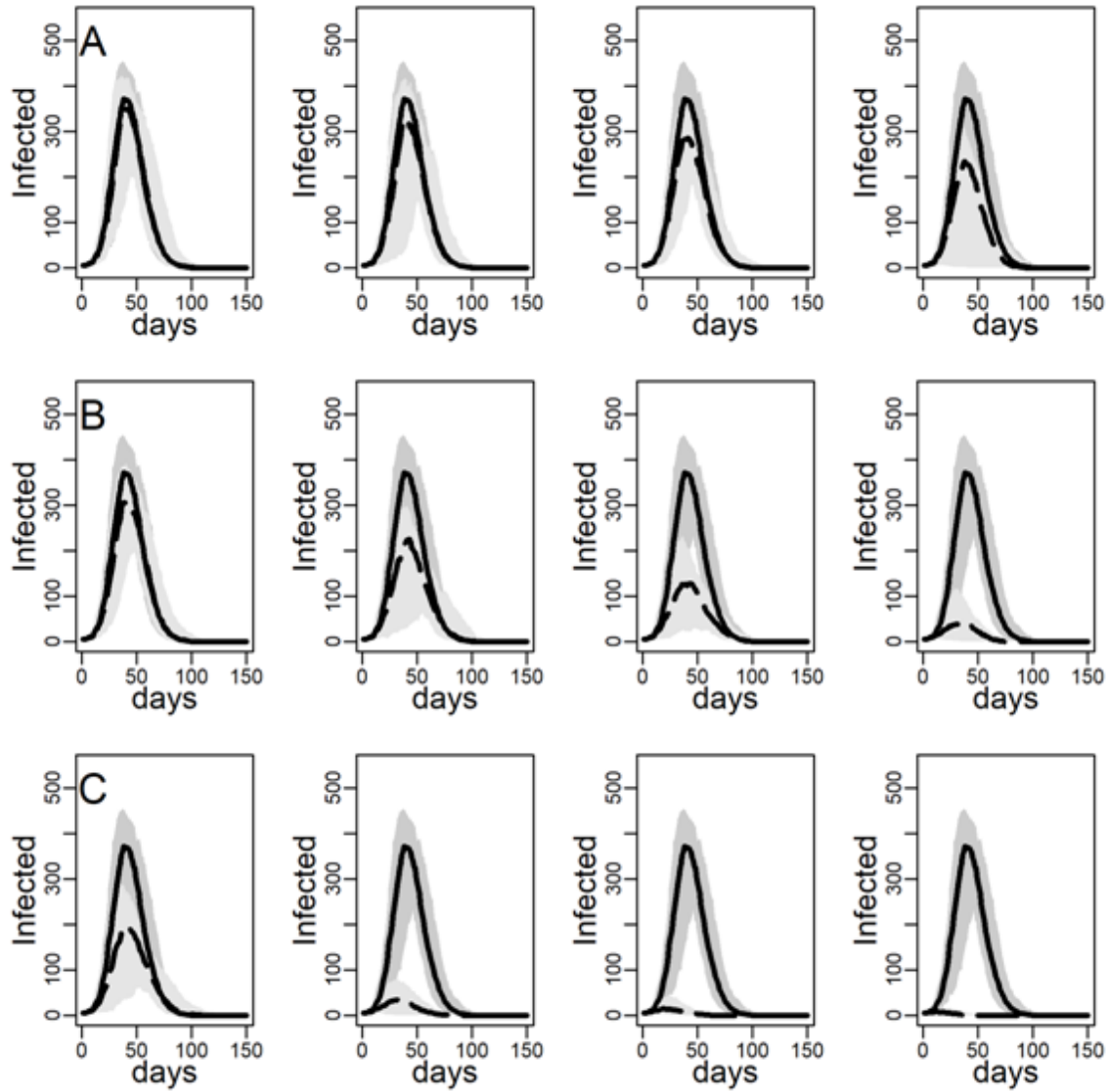

**Figure A3g.** Number of infected agents (y-axis) by days (x-axis) (median of 100 replications) under three different interventions (rows) targeting 1, 3, 5, or 10 agents per day (columns). A – NO-TARGET intervention; B – CONTACT-TARGET intervention. C – HUB-TARGET intervention. Lower and upper bounds of the shaded areas correspond to the 5[th] percentiles and 95[th] percentiles of the 100 replications. Solid line: *Degree-Calibrated* (DC) network; dashed line: interventions. Dyadic transmission probability (r=0.03) & Local clustering (Cc=0.50). *n* = 2,029 agents.

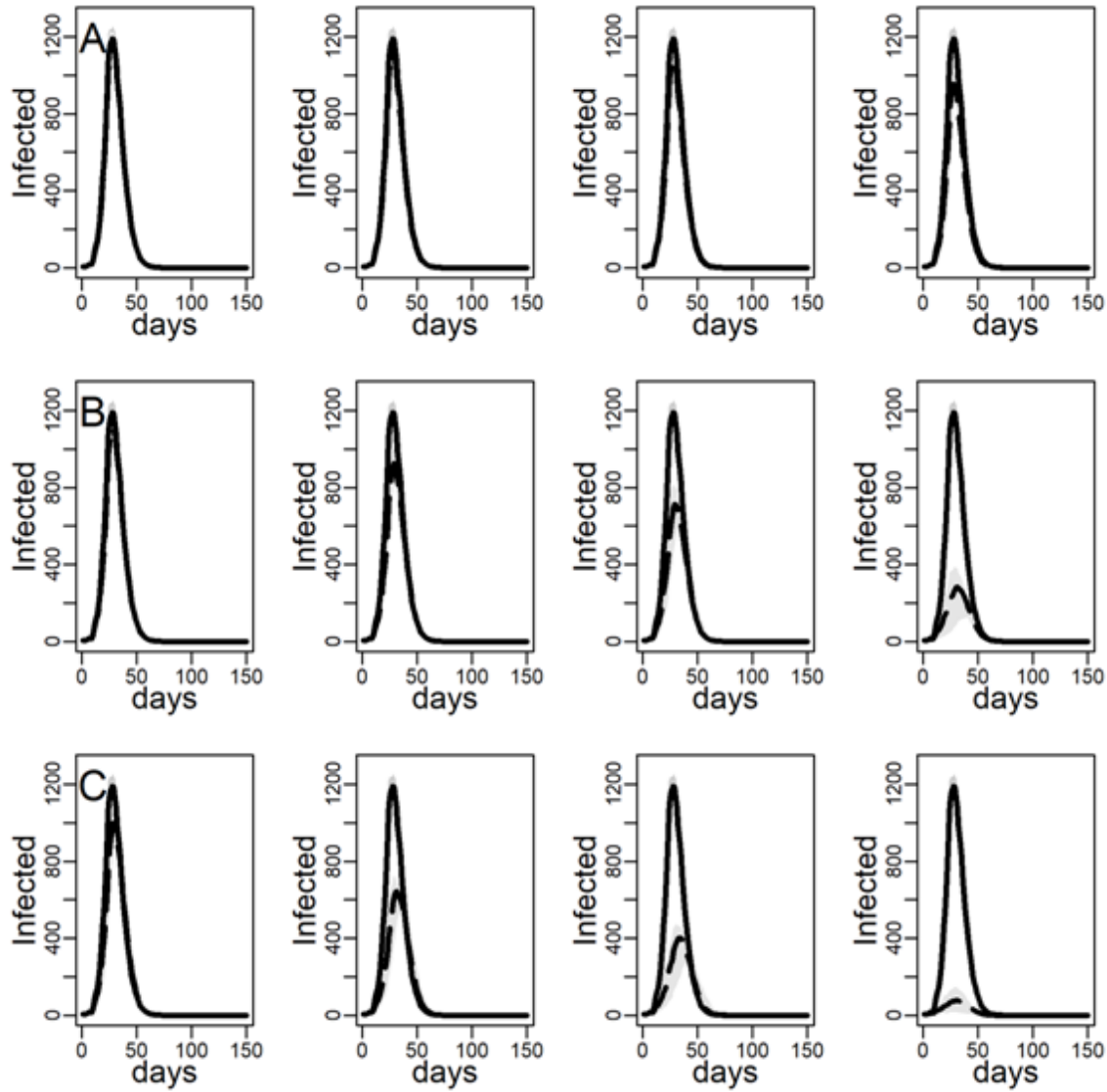

**Figure A3h.** Number of infected agents (y-axis) by days (x-axis) (median of 100 replications) under three different interventions (rows) targeting 1, 3, 5, or 10 agents per day (columns). A – NO-TARGET intervention; B – CONTACT-TARGET intervention. C – HUB-TARGET intervention. Lower and upper bounds of the shaded areas correspond to the 5th percentiles and 95th percentiles of the 100 replications. Solid line: *Degree-Calibrated* (DC) network; dashed line: interventions. Dyadic transmission probability (r=0.07) & Local clustering (Cc=0.04). *n* = 2,029 agents.

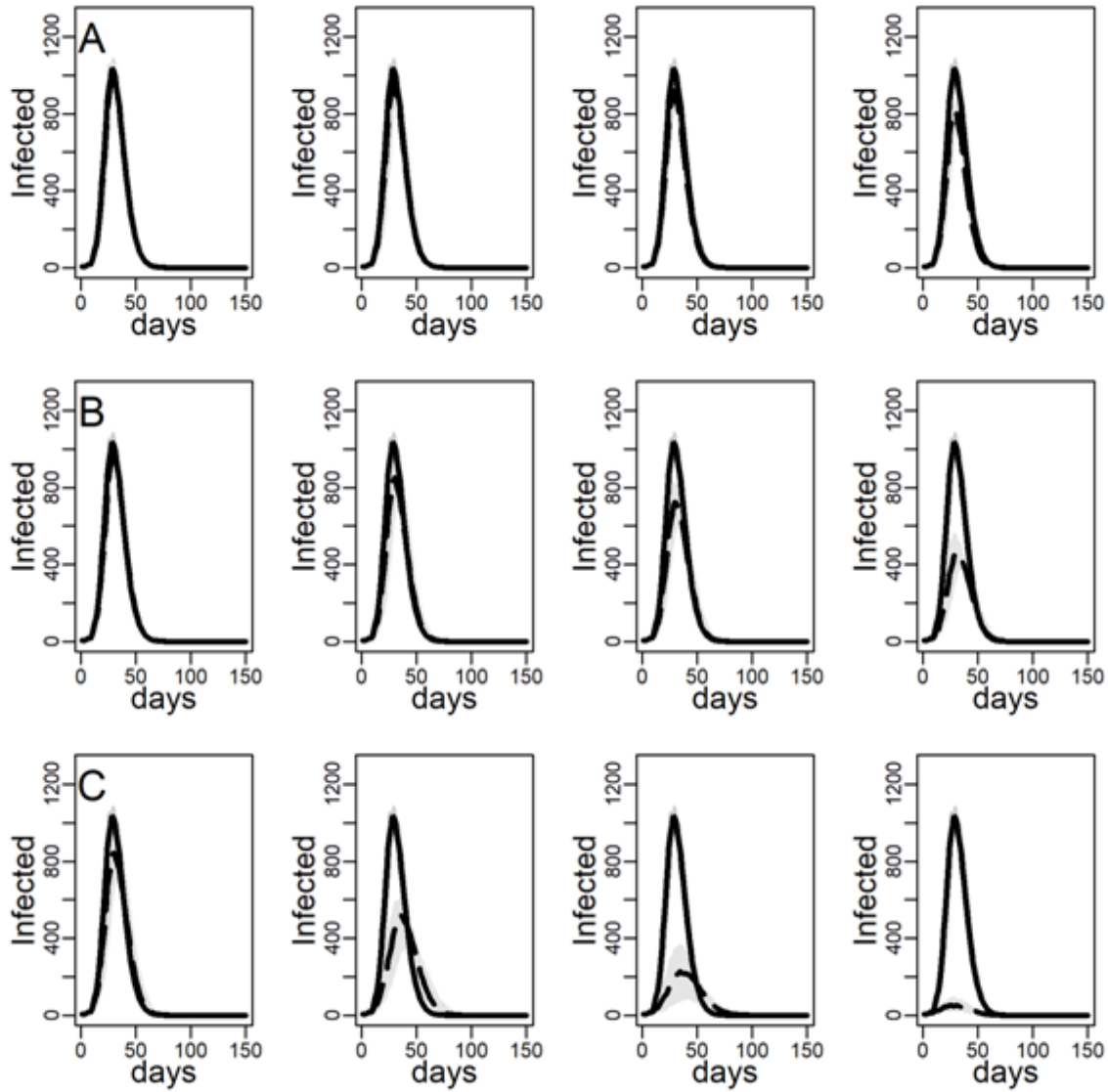

**Figure A3i.** Number of infected agents (y-axis) by days (x-axis) (median of 100 replications) under three different interventions (rows) targeting 1, 3, 5, or 10 agents per day (columns). A – NO-TARGET intervention; B – CONTACT-TARGET intervention. C – HUB-TARGET intervention. Lower and upper bounds of the shaded areas correspond to the 5$^{th}$ percentiles and 95$^{th}$ percentiles of the 100 replications. Solid line: *Degree-Calibrated* (DC) network; dashed line: interventions. Dyadic transmission probability (r=0.07) & Local clustering (Cc=0.42). *n* = 2,029 agents.

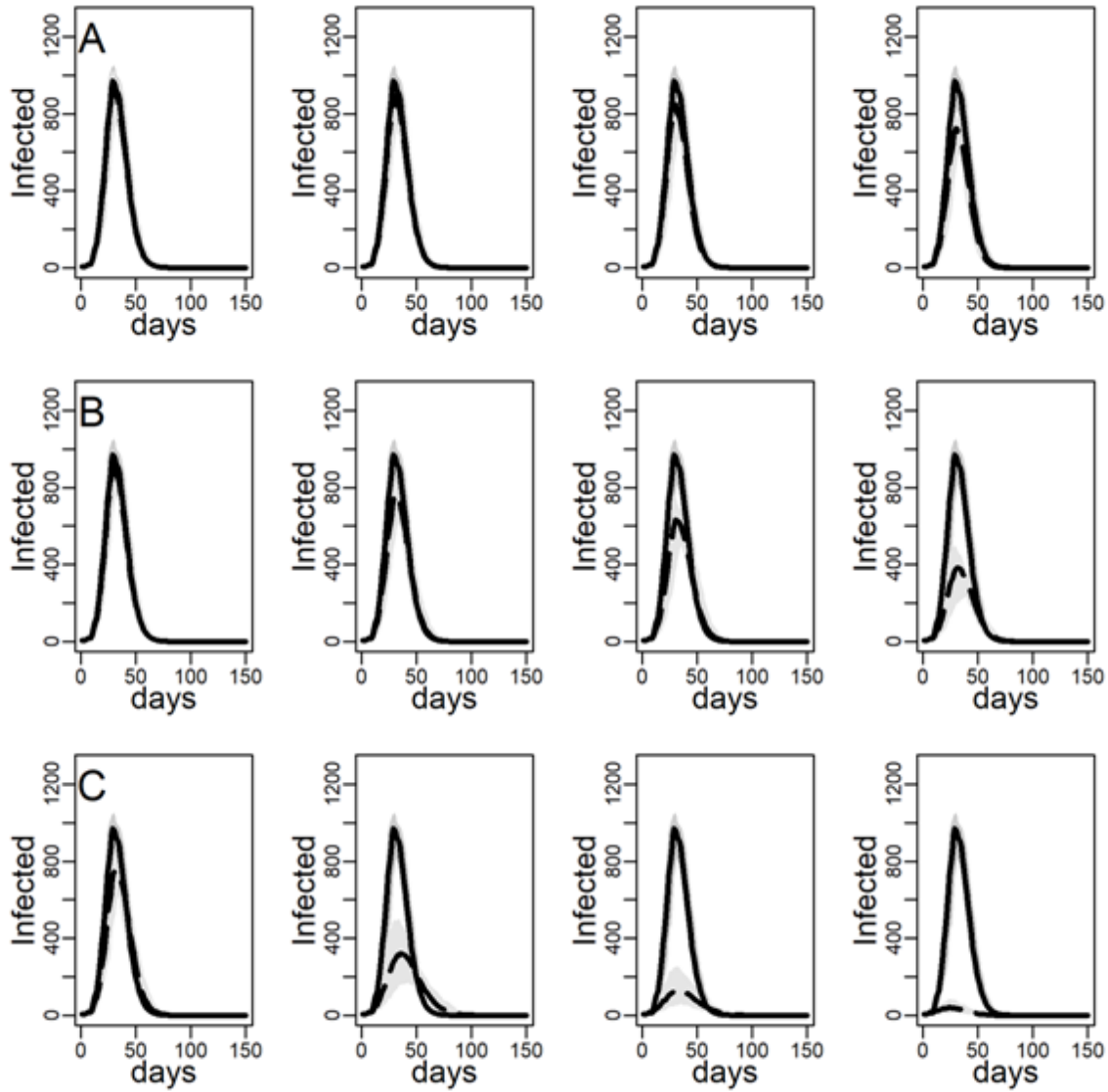

**Figure A3l.** Number of infected agents (y-axis) by days (x-axis) (median of 100 replications) under three different interventions (rows) targeting 1, 3, 5, or 10 agents per day (columns). A – NO-TARGET intervention; B – CONTACT-TARGET intervention. C – HUB-TARGET intervention. Lower and upper bounds of the shaded areas correspond to the 5[th] percentiles and 95[th] percentiles of the 100 replications. Solid line: *Degree-Calibrated* (DC) network; dashed line: interventions. Dyadic transmission probability (r=0.07) & Local clustering (Cc=0.50). *n* = 2,029 agents.